\documentclass[superscriptaddress,twocolumn,showpacs]{revtex4}
\usepackage[dvipdfmx]{graphicx}
\usepackage{color}
\usepackage{dcolumn}
\usepackage{amsmath}
\usepackage{dsfont}
\usepackage{epsfig}
\usepackage{soul}
\usepackage{bm}
\usepackage{amssymb}
\usepackage[dvipsnames]{xcolor}
\newcommand{\be}{\begin{equation}}
\newcommand{\ee}{\end{equation}}

\begin{document}

\title{Nucleation of transition waves via collisions of elastic vector solitons}

\author{H. Yasuda}
\affiliation{Department of Mechanical Engineering and Applied Mechanics, University of Pennsylvania, Philadelphia, Pennsylvania 19104, USA}
\affiliation{Aviation Technology Directorate, Japan Aerospace Exploration Agency, Mitaka, Tokyo 1810015, Japan}

\author{H. Shu}
\affiliation{Department of Mechanical Engineering and Applied Mechanics, University of Pennsylvania, Philadelphia, Pennsylvania 19104, USA}

\author{W. Jiao}
\affiliation{Department of Mechanical Engineering and Applied Mechanics, University of Pennsylvania, Philadelphia, Pennsylvania 19104, USA}

\author{V. Tournat}
\affiliation{Laboratoire d'Acoustique de l'Université du Mans (LAUM), UMR 6613, Institut d'Acoustique - Graduate School (IA-GS), CNRS, Le Mans Université, France }

\author{J. R. Raney}
\affiliation{Department of Mechanical Engineering and Applied Mechanics, University of Pennsylvania, Philadelphia, Pennsylvania 19104, USA}


\begin{abstract}
In this work, we show that collisions of one type of nonlinear wave can lead to generation of a different kind of nonlinear wave. Specifically, we demonstrate the formation of topological solitons (or transition waves) via collisions of elastic vector solitons, another type of nonlinear wave, in a multistable mechanical system
with coupling between translational and rotational degrees of freedom. 
We experimentally observe the nucleation of a phase transformation arising from colliding waves, 
and we numerically investigate head-on and overtaking collisions of solitary waves with vectorial properties (i.e., elastic vector solitons).
Unlike KdV-type solitons, which maintain their shape despite collisions, our system shows that collisions of two vector solitons can cause nucleation of a new phase via annihilation of the vector soltions, triggering the propagation of transition waves. The propagation of these depends both on the amount of energy carried by the vector solitons and on their respective rotational directions. 
The observation of the initiation of transition waves with collisions of vector solitons in multistable mechanical systems serves as an example of new fundamental nonlinear wave interactions, and could also prove useful in applications involving reconfigurable structures.
\end{abstract}

\maketitle

Nonlinear wave phenomena have long been of significant interest among researchers, both as an object of fundamental curiosity and for their potential in a variety of applications. Solitons, in particular, have long been studied in contexts ranging from the Fermi-Pasta-Ulam-Tsingou lattice and KdV equation~\cite{Fermi1955,Zabusky1965,Remoissenet1996} to granular media 
~\cite{Coste1997,Nesterenko2001,spadoni2010generation,Avalos2009}, soft matter (e.g., liquid crystals)~\cite{Manton2004,Shen2022}, etc. 
Generally, solitons are characterized as localized wave packets propagating with a constant speed and shape in the presence of nonlinearity and dispersion. Moreover, solitons keep their shape even if they collide with other solitons, analogous to the behavior of particles. 

More recently, flexible
mechanical metamaterials have been designed that derive their nonlinearity from large internal rotations~\cite{Deng_2021}, enabling fundamentally new nonlinear wave phenomena. 
For example, unlike solitons in classical 1D systems (e.g., granular crystals~\cite{Coste1997,Avalos2009}), a system of rotating squares, in which each element has translational and rotational degrees of freedom (DOF), can exhibit the propagation of vector solitons with coupling between these DOFs~\cite{Deng2017,Deng_2018,Deng_2021} (i.e., vectorial nature), cnoidal waves~\cite{Mo2019}, and ``sound bullets''~\cite{Deng2019}. 
The coupling of multiple DOFs also enables rich collision behavior, including collisions analogous to classical soliton collisions, but also repelling, destruction, etc.~\cite{Deng2019_collision}.

Moreover, multistable versions of these systems can support propagation of transition waves (or topological solitons), 
an effect that may be introduced via geometric constraints, permanent magnets, etc.~\cite{Raney2016,Yasuda2020,Jin2020,Ramakrishnan2020,Bilal_2017,Frazier2017,Hwang2018,Deng2019a}. Transition waves are propagating nonlinear wave fronts that sequentially switch the structural elements from one stable state to another~\cite{Raney2016,Deng_2021}. 
Recent research efforts on the observation, manipulation, and applications of transition waves in mechanical systems~\cite{Hwang2018,Zareei2020} provide a macroscopic analogy with the fundamental processes of dynamic phase transitions \cite{marigo2004initiation,truskinovsky2006quasicontinuum}, phase transformations in crystalline materials \cite{porter2009phase,james1986displacive}, and damage propagation in solids \cite{ren2011micro,ashby1990damage}. 

In this work, we 
demonstrate that colliding vector solitons can annihilate to nucleate a phase transition in multistable 
mechanical systems (i.e., the vector solitons cease to exist when the new phase is nucleated). The new phase can subsequently propagate outward in the form of a transition wave. 
While transition waves have been previously triggered at the edge of multistable mechanical metamaterials~\cite{Nadkarni2016,Raney2016,Jin2020,Korpas2021}, we demonstrate here that soliton collisions can effectively initiate transition waves at arbitrary locations in the structure (governed by the timing and amplitude of impulses at the ends of the structure). Our findings provide insight about the fundamental interactions between different types of nonlinear waves, and may in the future be relevant to novel engineering applications related to deployable structures, reconfigurable robots, etc.

\begin{figure}[htbp]
    \centerline{ \includegraphics[width=0.5\textwidth]{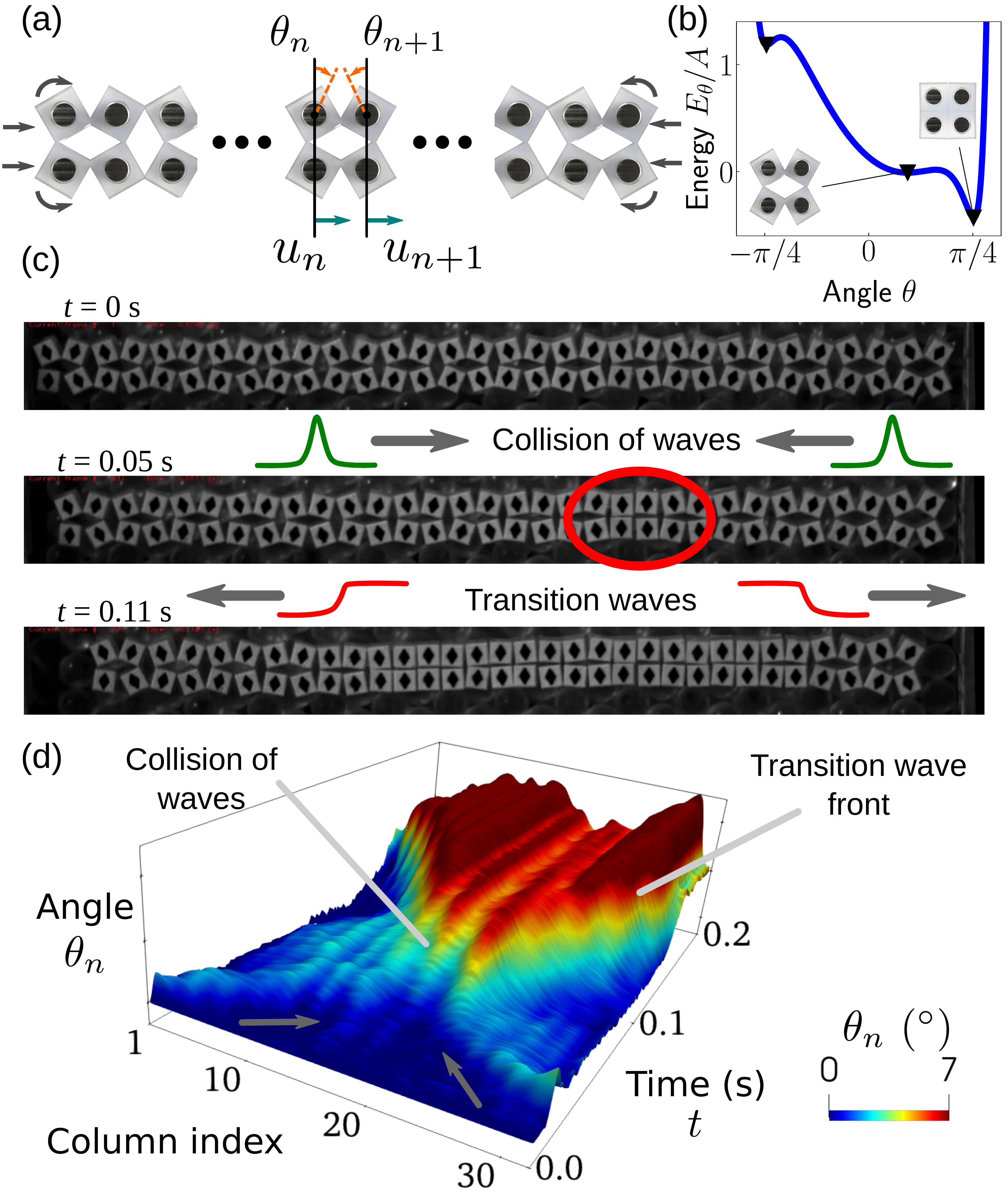}}
    \caption{
    (a) Schematic illustration of a chain of rotating squares with multistability. (b)~Multistable energy potential for a single complete unit of the chain (comprising two columns). (c)~Optical images taken with a high-speed camera at times $t= 0$~s, $0.05$~s, and $0.11$~s. (d)~Spatiotemporal plot extracted from the experimental video, showing the rotational angle $\theta_n$ for each column $n$ as a function of time. 
    }
    \label{fig:experiment}
\end{figure}

We start by experimentally exploring wave interactions in a multistable system by employing a 1D chain prototype composed of rotating squares and hinges (Fig.~\ref{fig:experiment}(a)). 
We define $u_n$ and $\theta_n$ as the relative displacement and rotational angle of the squares in the $n$th column (corresponding to the translational and rotational DOF), which are measured from their (initial) equilibrium state. 
Here, the initial equilibrium angle is expressed as $\theta^{(0)}$.
We 3D print a 32-column chain made of a silicone elastomer using direct ink writing~\cite{Shan2015}. To achieve multistable behavior, we embed a magnet in each square, resulting in an energy landscape with multiple equilibrium angles  (Fig.~\ref{fig:experiment}(b)).

To examine the wave interactions, we excite both sides of the chain simultaneously with two impactors. Figure~\ref{fig:experiment}(c) shows optical images obtained experimentally via a high-speed camera (Movie~1). We track the angles of each square ($\theta_n$) and plot the results as a function of the column index $n$ and time $t$, as shown in Fig.~\ref{fig:experiment}(d). 
Initially, the two pulses generated by the impactors propagate inward, toward the center of the chain (see the arrows in Fig.~\ref{fig:experiment}(d)). 
Then, once they meet in the middle of the chain, the collision induces larger rotations of the squares, leading to a phase transformation to another stable state (i.e., the system moves to a different energy minimum in the multistable energy landscape of Fig.~\ref{fig:experiment}(b)). 
Eventually, we observe a phase transformation that grows outward in the form of transition waves. 
While the propagation of transition waves in mechanical systems has been demonstrated previously, the initiation of these has been restricted to a direct excitation at the boundaries, and never (to our knowledge) via collisions of solitons. 
Our experimental observation suggests the possibility of controllable nucleation of transition waves at arbitrary locations within the system.

\begin{figure*}[htbp]
    \centerline{ \includegraphics[width=1.\textwidth]{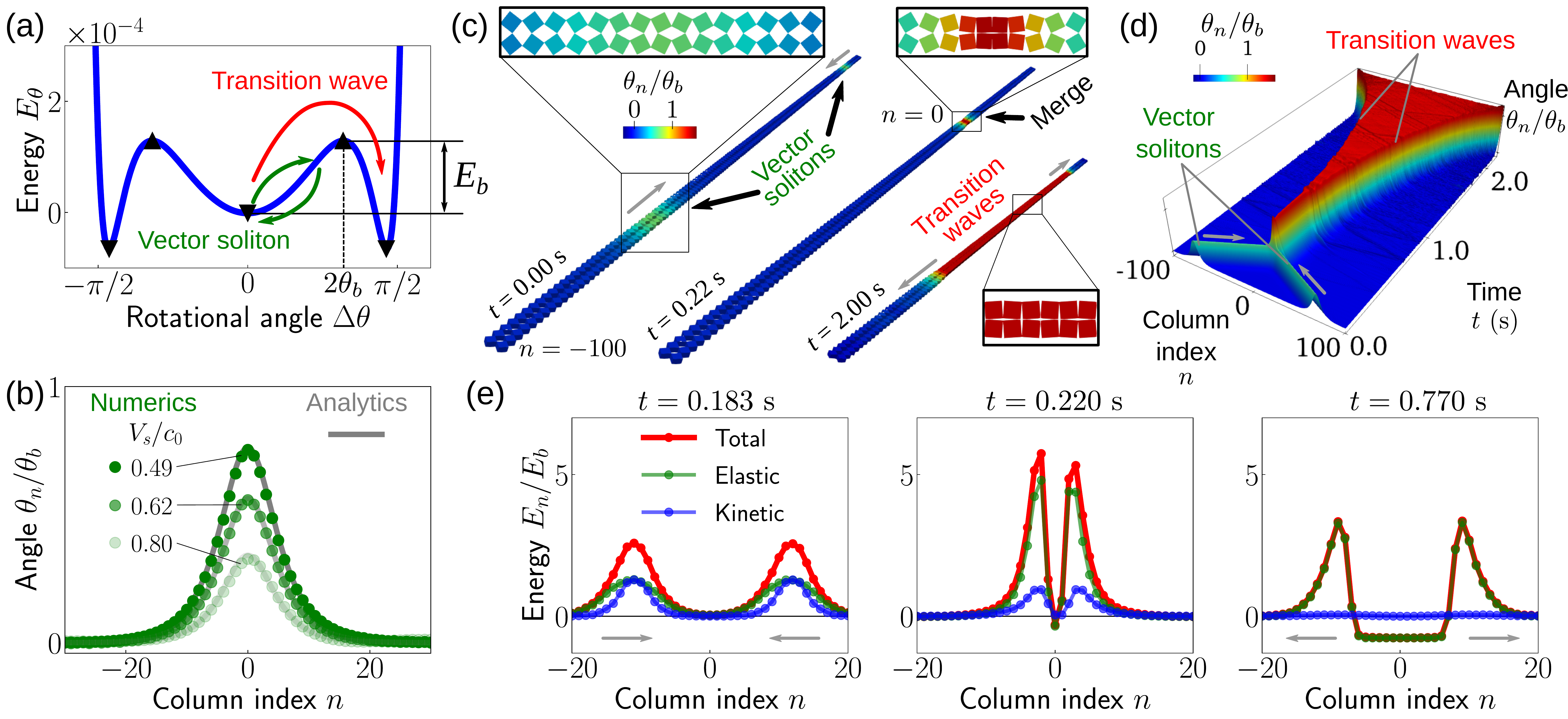}}
    \caption{(a) Multistable energy landscape. $E_b$ indicates the height of the energy barrier that the system needs to overcome to transition from 
    the center energy well to the right one. 
    (b) Angle profiles of solitary waves for three different wave speeds. (c) Snapshots of three-dimensional reconstructions of the numerical simulation at $t=0.00$, $0.22$, and $2.00$~s. The insets show the magnified views of a vector soliton propagating in the chain and the center part of the chain after the collision. (d)~The spatiotemporal evolution of the propagating waves, shown as the rotational angle of each column of squares as a function of time. (e)~Waveforms of colliding waves at i) $t=0.183$, ii) $0.220$, and iii) $0.770$ s, plotted as a function of energy and column index.} 
    \label{fig:Collision_solition}
\end{figure*}

To better understand the experimental observations, we introduce a numerical model and use it to investigate the nonlinear dynamics of colliding waves 
in this multistable system. 
First, to describe the interactions between neighboring squares, we model the hinges as a nonlinear spring element composed of axial and 
torsional springs, expressing the potential function of each hinge element as~\cite{Yasuda2020}:
\begin{equation} \label{eq:ResultantEnergy}
    U\left( \Delta \bm{l}, \Delta {{\theta }} \right) = \frac{1}{2}{{k}_{u}}{{\left| \Delta {\bm{l}} \right|}^{2}} + E_{\theta} \left( \Delta {{\theta }} \right),
\end{equation}
where $\Delta {\bm{l}}$ is the deformation of a linear axial spring with spring constant $k_u$,
and $\Delta \theta=2 \theta$
is the rotational angle of a 
torsional spring element with energy expressed by the Morse potential function ($V_{Morse}$) as: 
\begin{align} \label{eq:MorsePotential}
    & E_{\theta} \left( \Delta {{\theta }} \right) = \frac{1}{2}{{k}_{\theta }}{{\left( \Delta {{\theta }} \right) }^{2}}+{{V}_{Morse}}\left( \Delta {{\theta }} \right), \\
	& {{V}_{Morse}}\left( \Delta {{\theta }} \right) \nonumber \\
	& =A\left\{ {{e}^{2\alpha \left( \Delta {{\theta }}-2\theta _{M} \right)}}-2{{e}^{\alpha \left( \Delta {{\theta }}-2\theta _{M} \right)}} \right\} \nonumber \\
	& +A\left\{ {{e}^{-2\alpha \left( \Delta {{\theta }}-2\theta _{M} \right)}}-2{{e}^{-\alpha \left( \Delta {{\theta }_{n}}-2\theta _{M} \right)}} \right\}.
\end{align}
Here, $k_{\theta}$ is a linear spring constant. 
Also $A$ and $\alpha$ are parameters that alter the depth and width of the potential, respectively, and $\theta_{M}$ is a parameter that alters the equilibrium points. 
Figure~\ref{fig:Collision_solition}(a) shows the energy landscape of a single 
torsional spring element ($E_{\theta}$) as a function of a hinge rotational angle ($\Delta \theta=2\theta$) with three energy minima. Here, we denote the energy barrier and its location as $E_b$ and $\theta_b$.

We first consider the propagation of vector solitons in our multistable system by introducing several approximations and taking the continuum limit to derive the nonlinear Klein-Gordon equation. The wave forms of the derived solitary solutions are then compared with those of our numerical simulations. 
We approximate the potential energy landscape of the rotational spring element by using a 6th-order polynomial (the $\phi^6$ model). 
Then, by employing the traveling coordinate ($X = an-Vt$) and 
expressing the translational and rotational variables as $B_{u}(X)=u_n(t)/a$ and $B_{\theta}(X)=\theta_n(t)$, respectively, 
we obtain the following nonlinear Klein-Gordon equation for traveling waves: 
\begin{equation} \label{eq:NKG}
    \frac{\partial^2 B_{\theta}}{\partial X^2} = \eta_1 B_{\theta}-\eta_3 B_{\theta}^3 - \eta_5 B_{\theta}^5
\end{equation}
Note that a relationship between $B_u$ and $B_{\theta}$ (the subscripts $u$ and $\theta$ denote translational and rotational motions) holds for traveling waves, i.e., coupling behavior 
(Note~5 in the Supplemental Material \cite{SI}). 
This nonlinear Klein-Gordon equation supports a soliton solution~\cite{Khare2008}:
\begin{equation} \label{eq:soliton}
    B_{\theta} = \frac{A_{\theta} \text{sech} (\sqrt{\eta_1}X)}{\sqrt{1-D \tanh^2 (\sqrt{\eta_1}X)}},
\end{equation}
where $A_{\theta}^2=2\eta_1(1+D)/\eta_3$ and $(1+D)^2/D = 3\eta_3^2/(4\eta_1 \eta_5)$. 
Note that $\eta_1$, $\eta_3$, and $\eta_5$ are determined by the wave speed and axial/torsional spring constants, which alter both the amplitude $A_{\theta}$ and characteristic width $1/\sqrt{\eta_1}$.

Figure~\ref{fig:Collision_solition}(b) shows the rotational angle profiles obtained from the numerical simulation for a chain with 802 columns (green markers) and the soliton solution (solid line) for three different  
(normalized) wave speeds: $V_s/c_0=0.49$, 0.62, and 0.80. 
The soliton solution based on the potential energy $\phi^6$ approximation agrees well with the numerical simulations. Since the rotational DOF is coupled with the translational motion, this solitary wave 
possesses a vectorial nature.

Next, we study how multiple vector solitons interact with each other. To start, we consider the behavior of vector solitons colliding head-on. 
We inject two vector solitons with $V_s/c_0= \pm 0.56$ (i.e., two solitary waves propagating inward) at each end of a 
chain of 202 columns, and we solve the equations of motion using the fourth-order Runge-Kutta method. 
Figure~\ref{fig:Collision_solition}(c) shows numerically simulated spatiotemporal collision events in the chain (Movie~2). Figure~\ref{fig:Collision_solition}(d) plots the extracted rotation of each square, showing 
two vector solitons initially propagating inward with amplitudes smaller than 
$\theta_b$ (i.e., 
each remains in the first stable state) 
and annihilating once they collide at the center of the chain. Several columns show larger rotations than $\theta_b$ 
(i.e., $\theta_n/\theta_b>1$). The merging solitons thereby form a 
quasi-static disturbance, nucleating a phase transformation. 
Immediately after this, 
transition waves propagate outward from the site of collision, 
similar to the experimental observations (Fig.~\ref{fig:experiment}). Note, the propagation of such transition waves can be altered by tailoring the energy landscape~\cite{Yasuda2020}.

In addition to the rotational angle profiles, we analyze the collision behavior by considering the energy of each unit. 
Here, the total energy of the upper-half of the $n$th square element ($T_n$) is calculated as $T_n = K_n + E_n$, where $K_n$ is the kinetic energy
\begin{equation}  \label{eq:E_n}
    \begin{split}
    K_n &= \frac{1}{2}{{M}_{n}}\dot{u}_{n}^{2}+\frac{1}{2}{{J}_{n}}\dot{\theta }_{n}^{2},  \\
    E_n &=  {{U}_{n,1}}\left( \Delta {{\bm{l}}_{n,1}},\Delta {{\theta }_{n,1}} \right) + \frac{1}{2} {{U}_{n,2}}\left( \Delta {{\bm{l}}_{n,2}},\Delta {{\theta }_{n,2}} \right).
    \end{split}
\end{equation}
Note that for the right-most column, ${{U}_{n,1}}=0$. 
Figure~\ref{fig:Collision_solition}(e) shows the energy profiles consisting of total ($T_n$), elastic ($E_n$), and kinetic ($K_n$) energy components for three different time frames (Movie~3). 
Before the collision, the vector solitons have kinetic energy equivalent to their elastic energy. 
However, during the collision,  
the kinetic energy of several central units is fully transferred to elastic energy of neighboring units, such that the energy barrier is overcome and the corresponding units jump to another stable state. This triggers the formation of transition waves, in which the elastic energy becomes dominant.

\begin{figure}[htbp]
    \centerline{ \includegraphics[width=0.5\textwidth]{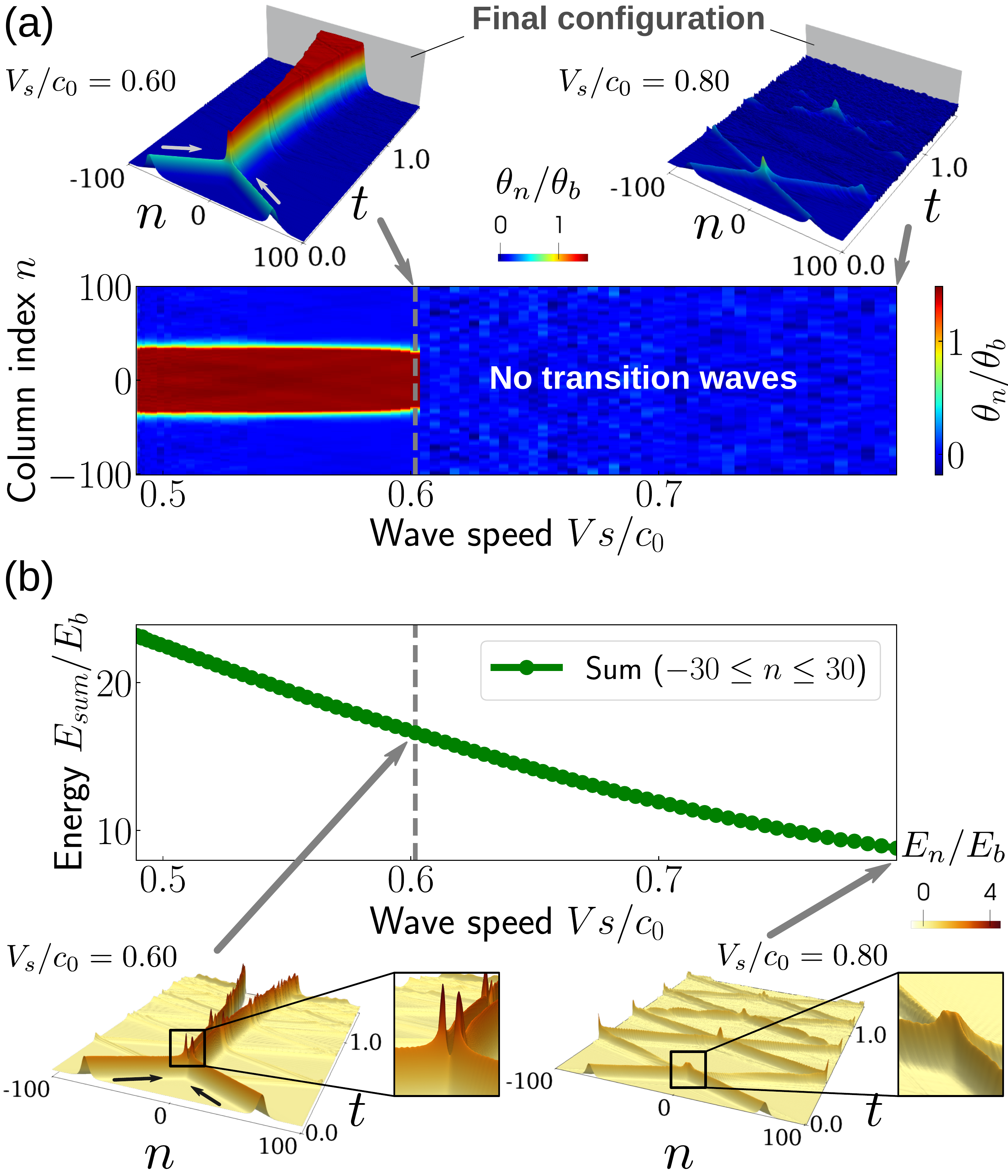}}
    \caption{(a) Collisions of vector solitons with various wave speeds. The surface plot represents the final configurations of the chain at $t=1.5$~s. The color denotes the normalized angle $\theta_n/\theta_b$ of each unit cell. (b) Summation of the total energy of a single vector soliton. The insets show the magnified view of the collision.} 
    \label{fig:threshold}
\end{figure}

Next, we analyze the process of transition wave nucleation by varying the amplitude (or equivalently the width or the wave speed) of colliding vector solitons. Note, due to the properties of the vector solitons in this system, specifically their subsonic nature, increasing the soliton velocity is equivalent to decreasing its amplitude, width, and energy. 
Numerical simulations are performed to examine the final chain configuration at $t=1.5$~s and to determine whether transition waves are formed or not. 
Results are plotted as a function of the wave speed $V_s/c_0$ of the two identical colliding solitons (Fig.~\ref{fig:threshold}(a)). 
Collisions of solitons with wave speeds $V_s/c_0 \lesssim 0.6$ form transition waves, as indicated by the red region in the figure, whereas collisions of faster vector solitons $V_s/c_0 \gtrsim 0.6$ (lower amplitude) do not trigger transition waves. 
Figure~\ref{fig:threshold}(b) shows the total energy carried by a single solitary wave for different wave speeds, which shows an energy threshold corresponding to a total soliton energy of $\sim 16.6 E_b$. 
The existence of this 
threshold indicates that while most of the kinetic energy is converted into elastic potential energy at the collision location, a minimum number of torsional springs within several units of the collision site need to be ``pushed'' over the potential barrier separating the stable states in order to induce nucleation.

\begin{figure*}[htbp]
    \centerline{ \includegraphics[width=1.\textwidth]{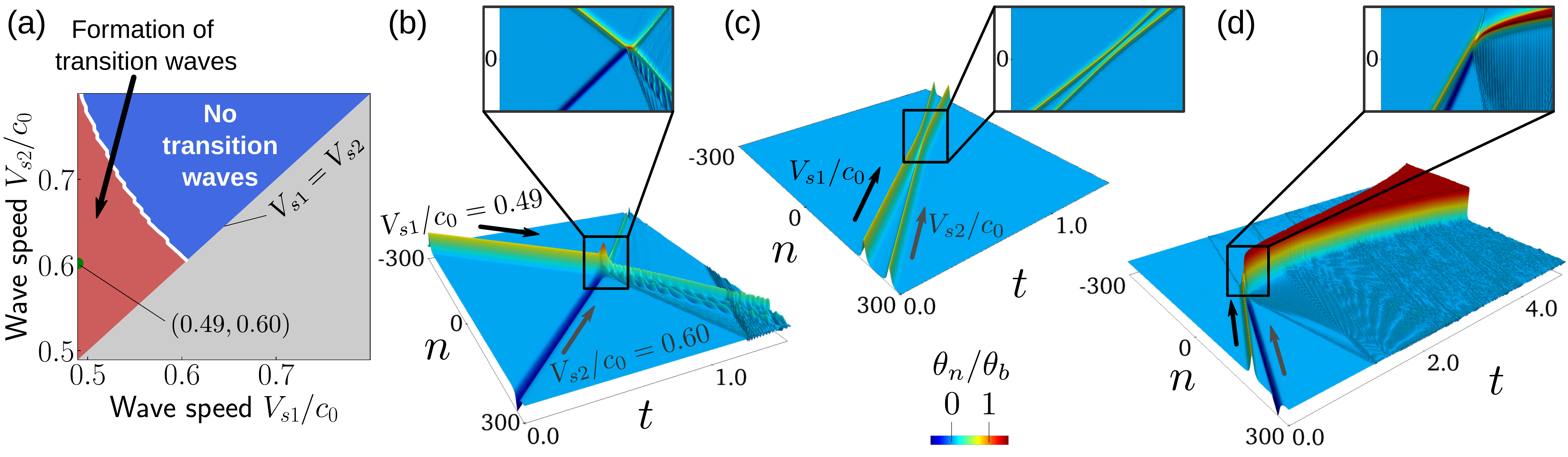}}
    \caption{(a) Collisions of vector solitons with distinct wave speeds ($V_{S1}\neq V_{S2}$, with $V_{S1}$ being the wave speed for the soliton propagating towards $n=300$ and $V_{S2}$ being the wave speed for the soliton propagating in the opposite direction). Due to symmetry, we investigate only the upper left region. (b)~Head-on collision of two vector solitons for the opposite rotation case. We also consider overtaking collisions for (c) same rotation and (d) opposite rotation.}
    \label{fig:overtaking}
\end{figure*}

The inset magnified views of the site of collision in Fig.~\ref{fig:threshold}(b) show split energy peaks instead of two waves merging into a single peak, which is a different picture from the superposition of linear waves or even from the classical soliton collision. These features arise from the multistable character of the metamaterial, allowing the colliding solitons to convert into quasi-static transition wave fronts, initially slow and carrying mostly potential elastic energy. Note that 
wave packets with smaller amplitudes (larger velocity) also emerge from the collision point, as do 
even smaller amplitude linear 
waves (Fig.~S8). 

We also numerically explore other collision scenarios, including head-on and overtaking collisions of vector solitons with distinct wave speeds/amplitudes and rotational directions (or chirality). 
Similar to Fig.~\ref{fig:threshold}(a), we examine the formation of transition waves from two vector solitons propagating from both ends of the chain, with the same chirality but with distinct wave speeds $V_{s1}\neq V_{s2}$ (and, hence, amplitudes). 
The results are shown in Fig.~\ref{fig:overtaking}(a), with the red and blue regions indicating that transition waves have or have not nucleated, respectively. 
Although two colliding solitary waves carry different amounts of energy before the collision, nucleation and propagation of transition waves can be observed (red region). In particular, nucleation takes place easily as we decrease the wave speed of (at least one of) the vector solitons (i.e., larger amplitude). 

We also consider different chiralities and propagation directions.
For example, vector solitons with opposite rotational directions can propagate simultaneously. 
The head-on collision of two vector solitons ($V_{s1}/c_0=0.49$ and $V_{s2}/c_0=0.60$) with the same rotational direction creates transition waves, as we discussed previously 
(Movie~4).
In contrast, if two solitons with opposite rotation collide, the one with larger energy breaks into several components that propagate outward instead of merging or repelling (Fig.~\ref{fig:overtaking}(b) and Movie~5). 
Note, 
if two solitons with opposite rotation propagate at identical (absolute) wave speed, they repel each other (see Fig.S10 in the Supplementary Materials \cite{SI}).

Figure~\ref{fig:overtaking}(c) shows the simulation results for the case in which two vector solitons propagate in the same direction, with the faster soliton ($V_{s2}/c_0=0.60$) catching up to the slower soliton. 
If the rotational directions of the two vector solitons are identical, the overtaking interaction only leads to a shift in phase (Movie~6), similar to classical KdV-type soliton collisions. 
However, if the vector solitons have opposite rotational directions, the overtaking interaction can lead to the formation of transition waves (Fig.~\ref{fig:overtaking}(d) and Movie~7), which is the opposite phenomenon than that observed for head-on collisions.

In conclusion, we have studied collisions of elastic vector solitons in a multistable mechanical metamaterial and the process by which these collisions can nucleate phase transitions. 
Our experimental observations and numerical analyses suggest that there is a threshold 
of vector soliton amplitude/energy, above which colliding vector solitons will cause nucleation and propagation of transition waves, and below which 
they will not. 
Using the validated numerical model, 
we have also explored different types of nonlinear wave interactions in the multistable system, such as head-on collisions of vector solitons possessing different amplitudes or chiralities (direction of rotation), or collisions of co-propagating solitons. 
Depending on the specific interaction, the results of these collisions can lead or not lead to conversion to transition waves, demonstrating the richness of this 
multistable mechanical platform for studying nonlinear waves. Furthermore, we note that this system shows signs of integrability for a range of soliton
amplitudes below the energy barrier (KdV soliton-like vector solitons with expected collisional
behavior) but also exhibits characteristics of non-integrability when the full tristable potential is
explored, as illustrated by the collision of solitons of sufficiently large amplitude, leading to the
nucleation of transition waves. Such a wide array of effects opens up new possibilities for fundamental research, by raising questions related to the transposition of these nonlinear processes into higher dimensions, into inhomogeneous or graded metamaterials, or for other nonlinear wave manifestations.
Also, our results may be relevant to new developments in topological metamaterials (such as twisted structures, graphene ~\cite{dai2016twisted}, etc.), where multistability and the ability to control morphology could lead to new approaches for controlling wave propagation.
One can, in addition, foresee possible future applications of this remote control of transition waves in, e.g., reconfigurable system design in robotics, space structures, medical devices, or in advanced materials for vibration damping, mechanical logic devices, and other emerging areas. 
\section*{Acknowledgements}
The authors gratefully acknowledge support via NSF award number 2041410, AFOSR award number FA9550-19-1-0285, and DARPA YFA award number W911NF2010278. 
H.Y. acknowledges the support of KAKENHI (22K14154). VT acknowledges support from project ExFLEM ANR-21-CE30-0003.
H.Y. and H.S. contributed equally to this work.



\bibliographystyle{unsrt}
\bibliography{References.bib}

\end{document}


\title{\textsf{Supplementary Information:\\Nucleation of transition waves via collisions of elastic vector solitons}}

\maketitle


\section*{Supplementary  Note 1: \textbf{Design and fabrication of experimental prototypes}}
\begin{figure}[htbp]
    \centerline{ \includegraphics[width=1\textwidth]{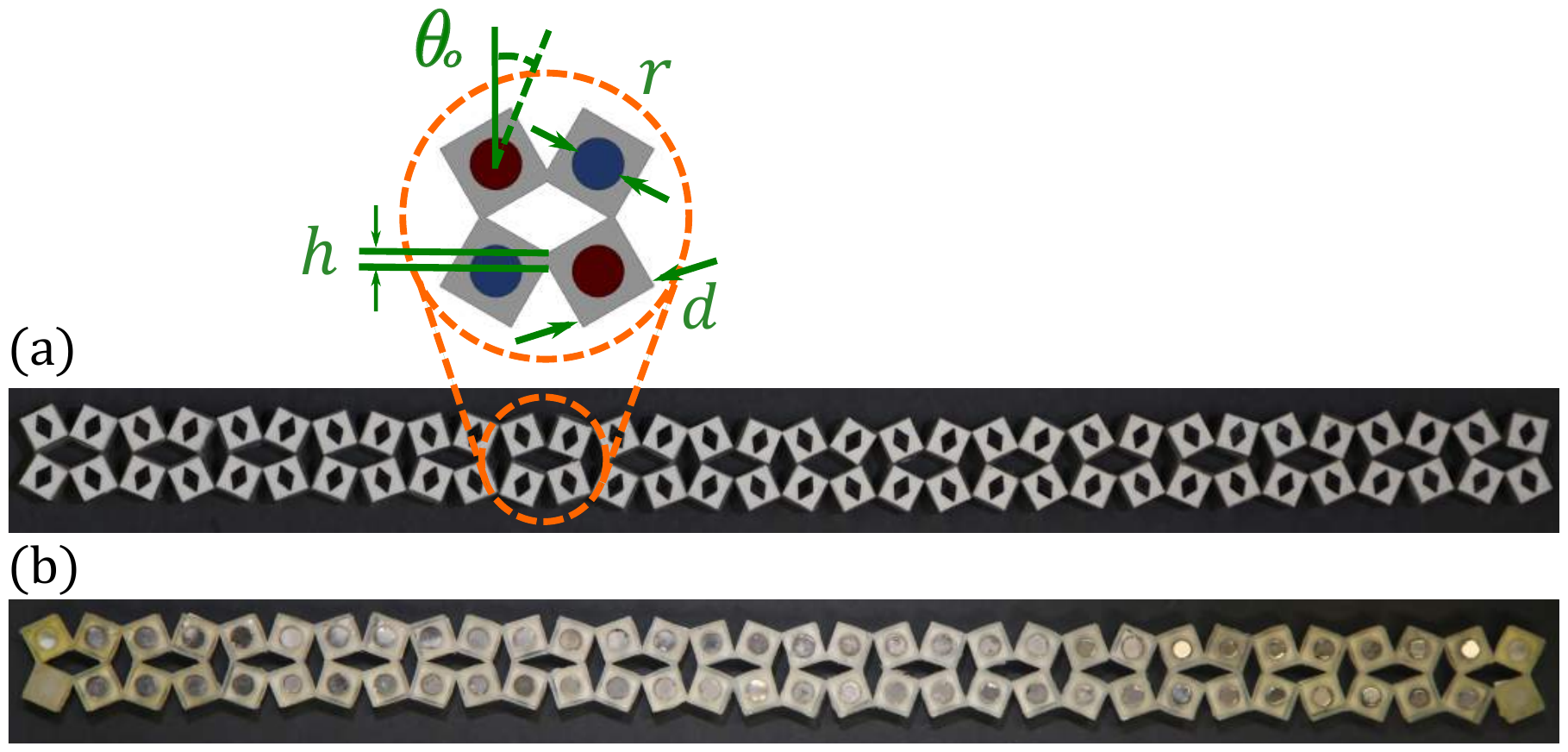}}
    \caption{(a). Optical image of the experimental specimen from the top, showing 32 columns of bistable rotating squares with permanent magnets embedded in their centers, and diamond-shaped trackers on the surface. (b)~The same specimen but from the bottom; the circles in the squares are the permanent magnets (no trackers are present on this side).} 
    \label{fig:SI_fabrication}
\end{figure}

In this work, we fabricated a 1D chain of bistable elements (Fig.~\ref{fig:SI_fabrication}), containing 32 columns of elastomeric rotating squares. The squares have edge length $d = 10$~mm and are rotated by an angle $\theta_0 = 15^{\circ}$ with respect to the vertical axis. Using direct ink writing~\cite{Raney15}, we first fabricate the hinges and outline of each square using a polydimethylsiloxane (PDMS) ink, as in \cite{Deng2017, Deng2019}, leaving a cylindrical space of radius $r = 6.35$~mm at the center. Note that the neighboring squares are connected via a thin ligament of PDMS, the length of which is controlled via the distance between the overlapping vertices, resulting in a hinge thickness $h = 1.2$~mm. After the initial pattern of the specimen is cured, additional material (Sylgard 184) is extruded to fill the space inside the border of each square (between the edge and the cylindrical void).  

Due to size constraints of the 3D printer, the 32-column sample was fabricated from two 16-column printed components, using the uncured printing material as glue. 
After curing the sample, permanent cylindrical magnets (D41-N52 Neodymium Magnets, K$\&$J Magnetics) are embedded at the center to provide attraction between adjacent squares. Note in Fig.\ref{fig:SI_fabrication} (b), magnets are removed from the edge columns to prevent the initial propagation of transition waves upon impact due to the boundary effect. Finally, 3D-printed (MakerGear M2, polylactic acid (PLA)), diamond-shaped trackers are adhered to the surface of each unit to allow tracking of the nodal rotation during dynamic testing. 

\clearpage
\section*{Supplementary Note 2: \textbf{Static testing}}
To characterize the static properties of the sample, we perform quasistatic tensile tests using an Instron model 68SC-5 in displacement control with a displacement rate of $0.02$~mm/s. Two aluminum fixtures are used to apply displacement to a specimen comprising four squares (two columns), as shown in Fig.~\ref{fig:SI_statictest}(a). 
Two set of tensile tests are conducted to obtain a complete set of force-displacement data: one set without magnets, and the other set with magnets. 

For the sample without magnets (Fig.\ref{fig:SI_statictest}(b)), we embed an aluminum rod at the center of each square. The two ends are aligned on the fixture's horizontal slot to allow free rotation and displacement of each square. We attach a 3D printed PLA onto the surface of each rotating square. The end of each rod is aligned in the slot.  Fig.~\ref{fig:SI_statictest}(b) and (c) indicate the locations of the applied force (green arrows) and the direction of rotation of each square (orange arrows). 
Figure~\ref{fig:SI_tensile} shows the force-displacement curve from the tensile test (blue line). 
As described below, we model our prototype as a structure composed of rigid squares and hinges (see also~\cite{Yasuda2020}). 
Based on the experimental results, we fit the parameters for the hinge components (red line in Fig.~\ref{fig:SI_tensile}) to extract spring constants $k_l$, $k_{\theta}$ and Morse potential parameters $A$, $\alpha$, $\theta_{M}$. 
We use this model to approximate the multistable energy landscape as a function of rotation angle, shown in Fig.~1b in the main text. 

\begin{figure}[!htbp]
    \centerline{ \includegraphics[width=0.6\textwidth]{Figures_SI/static testingv1.pdf}}
    \caption{ (a). Tensile test setup (Instron model 68SC-5 equipped with a custom aluminum fixture). (b-c).~Schematics of tensile tests without and with magnets, respectively. Green arrows indicate the direction of the applied force; orange arrows indicate the direction of rotation of each square.} 
    \label{fig:SI_statictest}
\end{figure}


\begin{figure}[htbp]
    \centerline{ \includegraphics[width=0.6\textwidth]{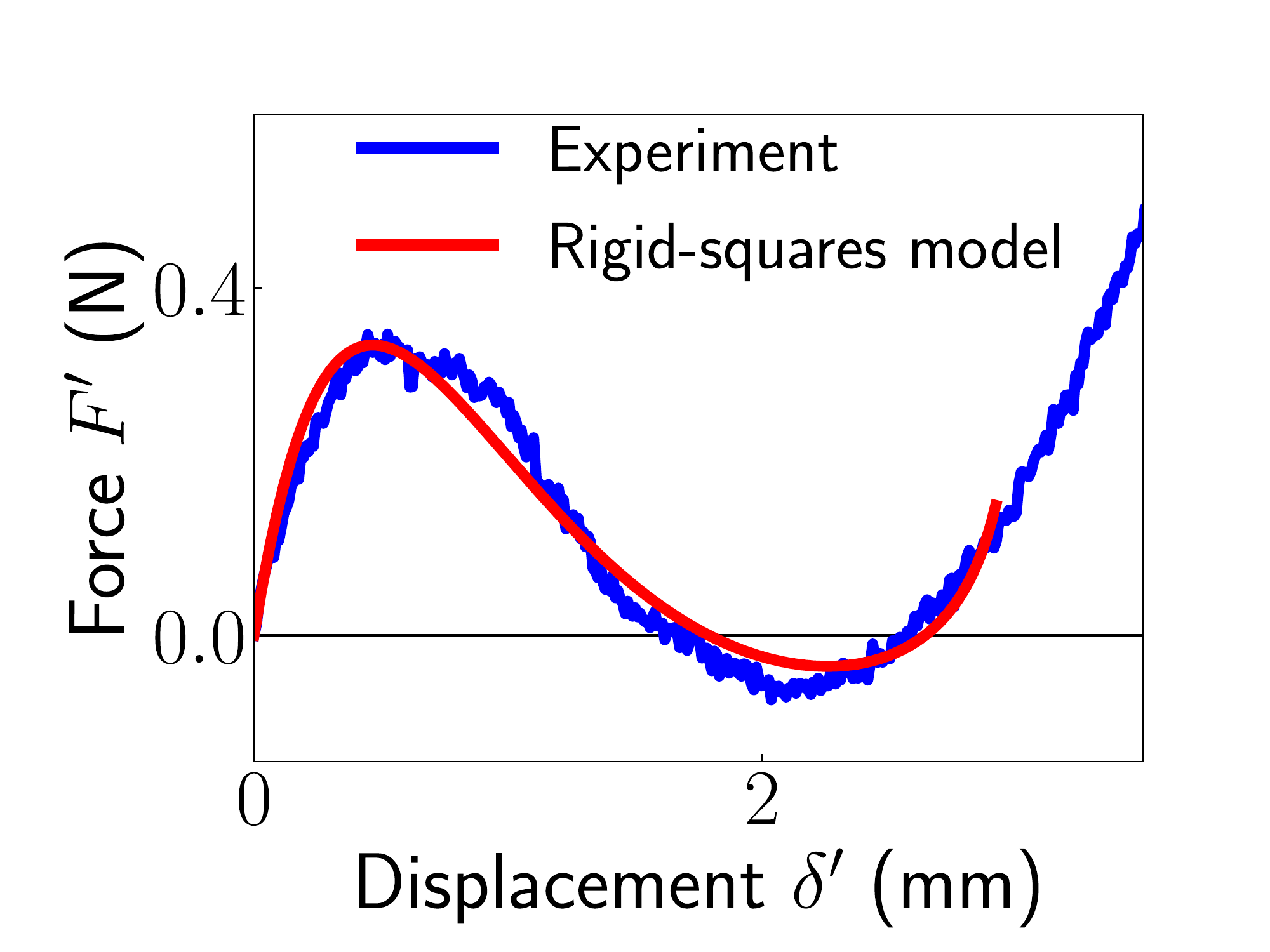}}
    \caption{Force-displacement relationship for the four-square structures. Blue and red lines indicate experimental and numerical results, respectively.} 
    \label{fig:SI_tensile}
\end{figure}

\clearpage

\newpage
\section*{Supplementary Note 3: \textbf{Dynamic testing}}
\label{SI:dynamics}
To dynamically test our experimental prototypes, we use a $45$~cm by $45$~cm surface filled with $15.9$~mm diameter acrylic balls to reduce friction (Fig.\ref{fig:SI_dynamictest}(a)). We use two custom aluminum impactors to simultaneously excite both sides of the sample. The excitation amplitude and velocity of the initial impulse are controlled via the impact distance (initial gap between the impactor and fixture) and the drive distance (initial gap between the impactor and sample), respectively. To study collisions of nonlinear waves and their subsequent effects, we use a high-speed camera (Photron FASTCAM Mini AX) and record the motion at 6400 frames per second. Diamond-shaped markers are adhered to the center of each square to allow tracking of the rotation and displacement of the squares, using a custom Python script (Fig.\ref{fig:SI_dynamictest}(b)). 
\begin{figure}[!htbp]
    \centerline{ \includegraphics[width=1\textwidth]{Figures_SI/dynamic testingv1.pdf} }
    \caption{ (a)~The experimental specimen rests on friction-reducing acrylic balls. Impactors excite the specimen at both ends. (b)~Top view of central four columns, with diamond-shaped markers adhered to the center of each square. (c)~Back view of central four columns (magnets are embedded inside).  (d)~View of squares from top and front.}
    \label{fig:SI_dynamictest}
\end{figure}


\newpage
\section*{Supplementary Note 4: \textbf{Numerical simulations}}
We have previously derived the equations of motion for a 1D chain of rotating squares~\cite{Yasuda2020}. The detailed derivations are summarized in this Note.

We assume the system comprises rigid rotating squares that are connected horizontally and vertically by hinges. 
We first model the elastic energy functions of two such hinge elements ($U_{n,1}$ for horizontal hinges, and $U_{n,2}$ for vertical connections) connected to the $n$th square by introducing the Morse potential function ($V_{Morse}$):
\begin{equation}
    {{U}_{n,m}}\left( \Delta {{\bm{l}}_{n,m}},\Delta {{\theta }_{n,m}} \right) = \frac{1}{2}{{k}_{u}}{{\left| \Delta {{\bm{l}}_{n,m}} \right|}^{2}}+\frac{1}{2}{{k}_{\theta }}{{\left( \Delta {{\theta }_{n,m}} \right)}^{2}} + {{V}_{Morse}}\left( \Delta {{\theta }_{n,m}} \right),
\end{equation}
where $m=1$ and $m=2$ indicate the horizontal and vertical components, respectively, $\Delta {\bm{l}}$ is the deformation of a linear axial spring with spring constant $k_u$, and $\Delta \theta=2 (\theta+\theta^{(0)}-\theta_{Lin})$ is the rotational angle of a torsional spring element. 
Here, $k_{\theta}$ is a spring constant for the linear part of the energy expression, and its equilibrium angle is denoted as $\theta_{Lin}$.
The elongation in the horizontal direction of a linear axial spring ($\Delta {\bm{l}}_{n,1}$) is expressed 
\begin{equation} \label{eq:Elongation}
    \Delta {{\bm{l}}_{n,1}} = \left[ \begin{matrix}
  {{u}_{n+1}}-{{u}_{n}}-l\cos \left( {{\theta }_{n+1}}+{{\theta }^{(0)}} \right)-l\cos \left( {{\theta }_{n}}+{{\theta }^{(0)}} \right)+2l\cos \left( {{\theta }^{(0)}} \right) \\ 
  -{{\left( -1 \right)}^{n+1}}l\sin \left( {{\theta }_{n+1}}+{{\theta }^{(0)}} \right)-{{\left( -1 \right)}^{n}}l\sin \left( {{\theta }_{n}}+{{\theta }^{(0)}} \right) \\ 
\end{matrix} \right]
\end{equation}
In this study, we neglect elongation in the vertical direction, i.e., $\Delta {{\bm{l}}_{n,2}}=\bm{O}$. 
For the rotational angle, we define
\begin{eqnarray} \label{eq:Elongation}
    \Delta {{\theta }_{n,1}} &=& {{\theta }_{n}}+{{\theta }_{n+1}} + 2(\theta^{(0)}-\theta_{Lin}), \\
    \Delta {{\theta }_{n,2}} &=& 2{{\theta }_{n}} + 2(\theta^{(0)}-\theta_{Lin}).
\end{eqnarray}
In addition, $V_{Morse}$ is composed of two Morse potential functions to represent the effect of the magnets, i.e.,
\begin{align} \label{eq:Morse_potential}
	{{V}_{Morse}}\left( \Delta {{\theta }} \right) & =A\left\{ {{e}^{2\alpha \left( \Delta {{\theta }}+2{{\theta }_{Lin}}-2\theta _{M} \right)}}-2{{e}^{\alpha \left( \Delta {{\theta }}+2{{\theta }_{Lin}}-2\theta _{M} \right)}} \right\} \nonumber \\
	& +A\left\{ {{e}^{-2\alpha \left( \Delta {{\theta }}+2{{\theta }_{Lin}}-2\theta _{M} \right)}}-2{{e}^{-\alpha \left( \Delta {{\theta }_{n}}+2{{\theta }_{Lin}}-2\theta _{M} \right)}} \right\},
\end{align}
where $A$ and $\alpha$ are parameters that alter the depth and width of the potential, respectively, and $\theta_{M}$ is a parameter that alters the equilibrium angles.

Next, we define the Hamiltonian of our multistable system as 
\begin{equation}  \label{eq:Hamiltonian}
    H = 2\sum\limits_{n=1}^{N}{\left\{ \frac{1}{2}{{M}_{n}}\dot{u}_{n}^{2}+\frac{1}{2}{{J}_{n}}\dot{\theta }_{n}^{2} \right\}} +2\sum\limits_{n=1}^{N-1}{{{U}_{n,1}}\left( \Delta {{\bm{l}}_{n,1}},\Delta {{\theta }_{n,1}} \right)} +\sum\limits_{n=1}^{N}{{{U}_{n,2}}\left( \Delta {{\bm{l}}_{n,2}},\Delta {{\theta }_{n,2}} \right)}
\end{equation}
where $M_n$ and $J_n$ are the mass and moment of inertia of the $n$th (single) square, respectively, and $u_n$ is the (relative) displacement.
Then, Hamilton's equations read
\begin{eqnarray}
    {{M}_{n}}{{\ddot{u}}_{n}} &=& -\frac{\partial H}{\partial {{u}_{n}}}, \\
    {{J}_{n}}{{\ddot{\theta }}_{n}} &=& -\frac{\partial H}{\partial {{\theta }_{n}}}.
\end{eqnarray}
By taking a derivative of the Morse potential function, we obtain
\begin{equation}  \label{eq:Hamiltonian}
	\begin{split}
    {{T}_{Morse}}\left( \Delta {{\theta }_{n,m}} \right) = &\frac{\partial {{V}_{Morse}}\left( \Delta {{\theta }_{n,m}} \right)}{\partial {{\theta }_{n}}} \\
    = &2\alpha A\left\{ {{e}^{2\alpha \left( \Delta {{\theta }_{n,m}}+2{{\theta }_{Lin}}-2\theta _{Morse} \right)}}-{{e}^{\alpha \left( \Delta {{\theta }_{n,m}}+2{{\theta }_{Lin}}-2\theta _{Morse} \right)}} \right\} \\
    - &2\alpha A\left\{ {{e}^{-2\alpha \left( \Delta {{\theta }_{n,m}}+2{{\theta }_{Lin}}-2\theta _{Morse} \right)}}-{{e}^{-\alpha \left( \Delta {{\theta }_{n,m}}+2{{\theta }_{Lin}}-\Delta \theta _{Morse} \right)}} \right\}.
    \end{split}
\end{equation}
Therefore, we get the equations of motion of the $n$th square as follows:
\begin{eqnarray}  \label{eq:EqMo}
    {{M}_{n}}{{\ddot{u}}_{n}} &=& {{k}_{u}}\left( {{u}_{n+1}}+{{u}_{n-1}}-2{{u}_{n}} \right) \nonumber \\
    &+& {{k}_{u}}l\left\{ \cos \left( {{\theta }_{n-1}}+{{\theta }^{(0)}} \right)-\cos \left( {{\theta }_{n+1}}+{{\theta }^{(0)}} \right) \right\} \\
    {{J}_{n}}{{\ddot{\theta }}_{n}} &=& -{{k}_{\theta }}\left( {{\theta }_{n+1}}+{{\theta }_{n-1}}+4{{\theta }_{n}}+6{{\theta }^{(0)}}-6\theta _{Lin} \right) \nonumber \\
    &-& {{k}_{u}}l\sin \left( {{\theta }_{n}}+{{\theta }^{(0)}} \right)\left( {{u}_{n+1}}-{{u}_{n-1}} \right) \nonumber \\
    &+& {{k}_{u}}{{l}^{2}}\cos \left( {{\theta }_{n}}+{{\theta }^{(0)}} \right) \nonumber \\
    && \left\{ \sin \left( {{\theta }_{n+1}}+{{\theta }^{(0)}} \right)+\sin \left( {{\theta }_{n-1}}+{{\theta }^{(0)}} \right) \right\} \nonumber \\
    &-& {{T}_{Morse}}\left( \Delta {{\theta }_{n,1}} \right)-{{T}_{Morse}}\left( \Delta {{\theta }_{n-1,1}} \right) \nonumber \\
    &-& {{T}_{Morse}}\left( \Delta {{\theta }_{n,2}} \right)
\end{eqnarray}

By using the equations of motion that we derived above, we perform numerical simulations to analyze the propagation of vector solitons.
The parameters used in this study are based on Ref.~\cite{Yasuda2020}, and are summarized in Table~\ref{tab:parameters}.
We use a longer chain composed of 802 columns.
To generate a vector soliton, we apply an impact to the chain, specifically setting the initial velocities of the left-most squares ($n=-400$) as $\left( \dot u_1, \dot \theta_1 \right) = \left( 2.34, 125 \right)$. 
Then, we solve the equations of motion by employing the Runge-Kutta method.

\begin{figure*}[htbp]
    \centerline{ \includegraphics[width=1.1\textwidth]{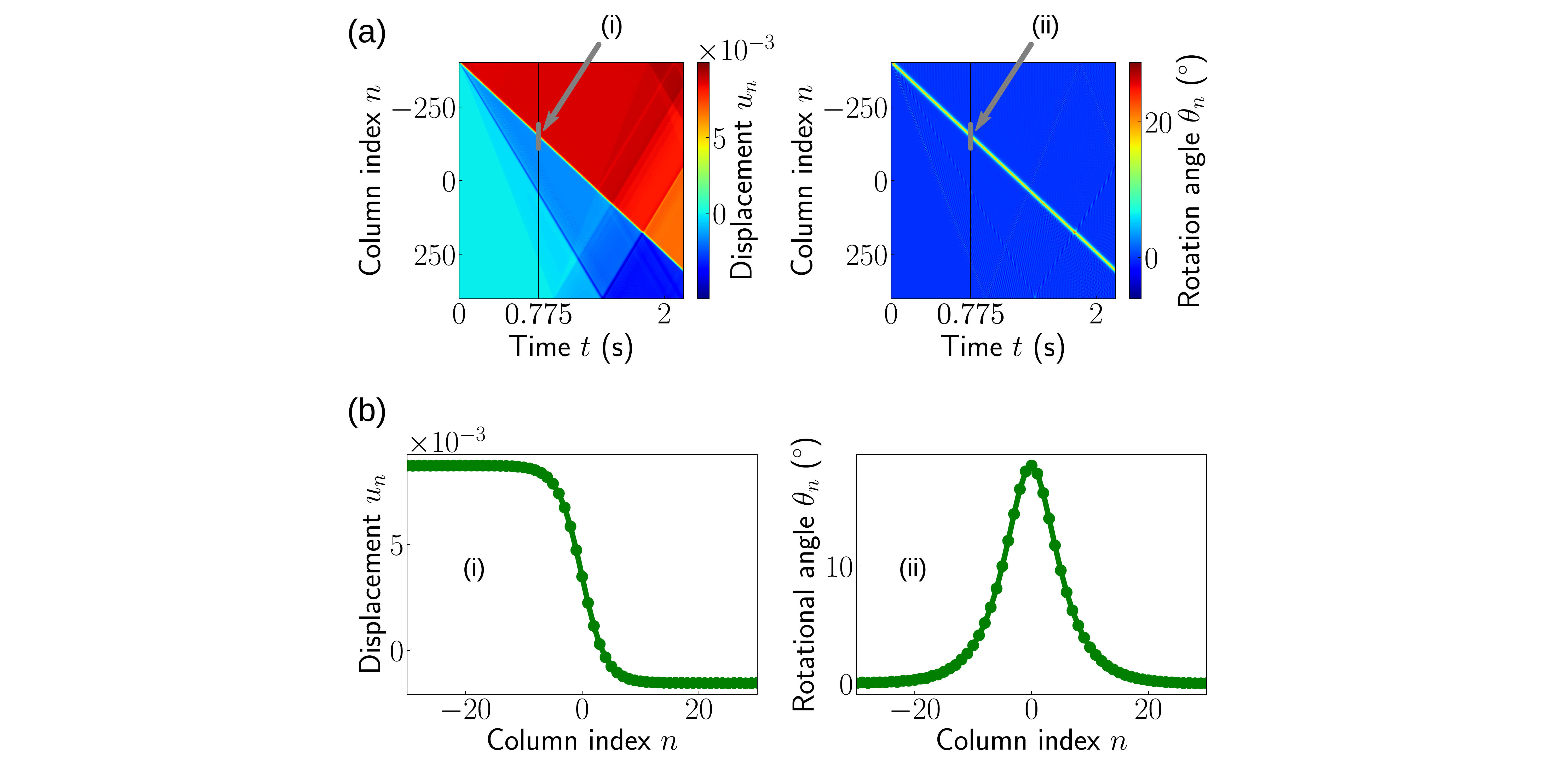}}
    \caption{(a) Space-time contour plots of wave propagation for displacement ($u_n$) and rotational angle ($\theta_n$). The initial condition of the left-most column ($n=-400$) is $\left( \dot u_1, \dot \theta_1 \right) = \left( 2.34, 125 \right)$. (b) We extract the wave form of a vector soliton at $t=0.775$ s (see the arrows (i) and (ii) in (a)). The wave forms are shifted along the horizontal axis so that the peak of the vector soliton is located at $n=0$.
    } 
    \label{fig:sim_longchain}
\end{figure*}

Figure~\ref{fig:sim_longchain} (a) shows the spatio-temporal plots of the displacement ($u_n$) and rotational angle ($\theta_n$) from the simulation. 
We observe the propagation of a single vector soliton with coupled translational and rotational motion. 
The wave form of the generated vector soliton at $t=0.775$~s is shown in  Fig.~\ref{fig:sim_longchain}(b), in which panels (i) and (ii) show the displacement and rotational angle profiles of the vector soliton (see also the grey colored arrows in Fig.~\ref{fig:sim_longchain} (a)).
We compare the extracted wave form from the simulation with the single-soliton solution from the nonlinear Klein-Gordon equation in Note~5. 
Also, we use the extracted wave profiles from longer-chain simulations for the collision analysis (e.g., see Fig.~\ref{fig:odd_numbered} for the numerical analysis to examine the colliding solitary waves in 202-column (even-numbered) and 203-column (odd-numbered) chains ). 
We employ this procedure to minimize the effect of small-amplitude linear waves created by the impact input.

\begin{table}[htbp]
\caption{Numerical constants used in the numerical and theoretical analysis.}\label{tab:parameters}
\begin{tabular}{cc}
\hline
Parameters        &                        \\ \hline
$\theta^{(0)}$   & 0                        \\
$\theta_{Lin}$   & 0                        \\
$k_u$   & 670.8  N/m                        \\
$k_{\theta}$ & $4.433 \times 10^{-4}$ N m                         \\
$A$  & $5.477 \times 10^{-4}$                \\
$\alpha$       & 4.966  \\
$\theta_{Morse}$         & 0.7442      \\
$M_n$  & $1.989 \times 10^{-3}$ g   \\
$J_n$  & $46.60 \times 10^{-9}$ kg m$^2$ \\
\hline
\end{tabular}
\end{table}

\begin{figure*}[htbp]
    \centerline{ \includegraphics[width=1.0\textwidth]{Figures_SI/odd_numbered_chain.pdf}}
    \caption{(a) Surface plot of energy for the even-numbered-column chain composed of 202 columns. (b)~Wave forms of vector solitons at $t=0.180$~s, 0.220~s, and 0.77~s. The numerical analysis for the odd-numbered-column chain (203 columns) is also shown in panels (c) and (d). For both cases, the wave speed of the vector solitons is $V_s/c_0=0.56$.
    } 
    \label{fig:odd_numbered}
\end{figure*}

\clearpage
\section*{Supplementary  Note 5: \textbf{Nonlinear Klein-Gordon equation}}
We also explore an analytical approach to study nonlinear wave dynamics of our multistable system, specifically vector solitons. 
We employ the nonlinear Klein-Gordon equation based on the $\phi^6$ potential to derive soliton solutions.
In this Note, we consider the case of $\theta^{(0)}=\theta_{Lin}=0$ for the sake of simplicity \cite{Yasuda2020}.
To analyze the system analytically, we derive a continuum model by taking a continuum limit.

\begin{figure*}[htbp]
    \centerline{ \includegraphics[width=0.8\textwidth]{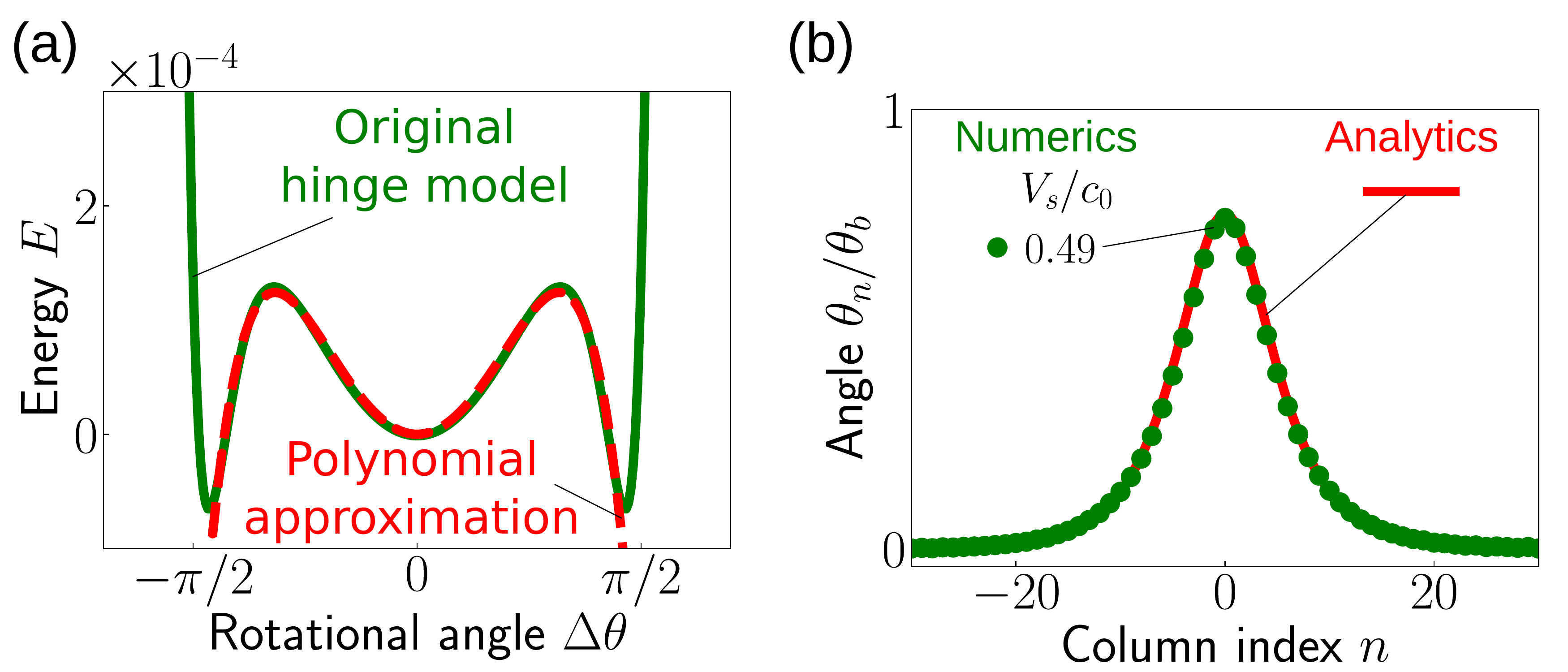}}
    \caption{(a) Energy landscape. (b) Wave profile obtained from the numerical simulation and single-soliton solution from the nonlinear Klein-Gordon equation based on the $\phi^6$ model.
    } 
    \label{fig:phi6}
\end{figure*}

First, we approximate the potential energy function of the hinge, particularly a torsional component ($E_{\theta}$), by using the following polynomial function:
\begin{equation}  \label{eq:E_poly}
    \begin{split}
    {{E}_{\theta }}\left( \Delta \theta  \right) &= \frac{{k}_{\theta }}{2}{{\left( \Delta \theta  \right)}^{2}} +{{V}_{Morse}}\left( \Delta \theta  \right) \nonumber \\
    &\approx \frac{{k}_{\theta }}{2}{{\left( \Delta \theta  \right)}^{2}}+\frac{{{C}_{4}}}{4}{{\left( \Delta \theta  \right)}^{4}}+\frac{{{C}_{6}}}{6}{{\left( \Delta \theta  \right)}^{6}},
    \end{split}
\end{equation}
To obtain the coefficients $C_4$ and $C_6$, the original energy curve $E_{\theta}$ is fitted by the polynomial function numerically, which results in $\left( C_4, C_6 \right) = \left( -2.73 \times 10^{-4}, -1.76 \times 10^{-4} \right)$.
Figure~\ref{fig:phi6}(a) shows the comparison between the original and fitted energy curves.
The curve fitting agrees well with the original energy function ($E_{\theta}$), especially in the center well where vector solitons can be formed.

By simplifying the trigonometric functions, i.e., $\sin \theta = \theta -{\theta }^{3} / {3!}$ and $\cos \theta = 1-{\theta }^{2}/{2!}$, we rewrite the equations of motion as follows:
\begin{eqnarray} \label{eq:EqMo_simplyfied}
    \frac{{{a}^{2}}}{c_{0,u}^{2}}{{\ddot{u}}_{n}} &=& \left( {{u}_{n+1}}+{{u}_{n-1}}-2{{u}_{n}} \right)+\frac{1}{4}a\left( \theta _{n+1}^{2}-\theta _{n-1}^{2} \right) \nonumber \\
    \frac{{{a}^{2}}}{c_{0,u}^{2}{{\gamma }^{2}}}{{{\ddot{\theta }}}_{n}} &=& -{{K}_{\theta }}\left( {{\theta }_{n-1}}+{{\theta }_{n+1}}+4{{\theta }_{n}} \right) -\frac{2}{a}{{\theta }_{n}}\left( {{u}_{n+1}}-{{u}_{n-1}} \right) \nonumber  \\ 
 &&  +\left( -\frac{{{\theta }_{n}}\theta _{n+1}^{2}}{2}-\frac{{{\theta }_{n}}\theta _{n-1}^{2}}{2}-2{{\theta }_{n}}+\frac{\theta _{n}^{3}}{3} \right) \nonumber  \\ 
 &&  +\left( {{\theta }_{n+1}}+{{\theta }_{n-1}}-\frac{\theta _{n+1}^{3}}{6}-\frac{\theta _{n-1}^{3}}{6}-\frac{\theta _{n}^{2}{{\theta }_{n+1}}}{2}-\frac{\theta _{n}^{2}{{\theta }_{n-1}}}{2} \right)  \nonumber  \\ 
 &&  -{{C'}_{4}}\left\{ {{\left( {{\theta }_{n}}+{{\theta }_{n+1}} \right)}^{3}}+{{\left( {{\theta }_{n-1}}+{{\theta }_{n}} \right)}^{3}}+8\theta _{n}^{3} \right\}  \nonumber  \\ 
 &&  -{{C'}_{6}}\left\{ {{\left( {{\theta }_{n}}+{{\theta }_{n+1}} \right)}^{5}}+{{\left( {{\theta }_{n-1}}+{{\theta }_{n}} \right)}^{5}}+32\theta _{n}^{5} \right\}
\end{eqnarray}
where we redefine the coefficients as
\begin{eqnarray}
    {{c}_{0,u}} &=& a\sqrt{\frac{{{k}_{u}}}{{{M}_{n}}}}, \\
    \gamma &=& a\sqrt{\frac{{{M}_{n}}}{4{{J}_{n}}}}, \\
    a &=& 2l, \\
    {{K}_{\theta }} &=& \frac{4{{k}_{\theta }}}{{{k}_{u}}{{a}^{2}}}, \\
    {{{C}'}_{i}} &=& \frac{4}{{{a}^{2}}{{k}_{u}}}{{C}_{i}} \ \ \ \ \  (i = 4,6).
\end{eqnarray}

 To derive a continuum model, we introduce the traveling coordinate $X=a n - V t$, and we rewrite the displacement and rotational angle as
\begin{eqnarray}
    {{u}_{n}}\left( t \right) &=& a{{B}_{u}}\left( X \right) \\
    {{\theta }_{n}}\left( t \right) &=& {{B}_{\theta }}\left( X \right).
\end{eqnarray}
Applying a Taylor series expansion to the displacement and rotational angle gives 
\begin{equation}  \label{eq:Taylor}
    B\left( X\pm a \right)=B\left( X \right)\pm a\frac{\partial B}{\partial X}+\frac{{{a}^{2}}}{2}\frac{{{\partial }^{2}}B}{\partial {{X}^{2}}}\pm \frac{{{a}^{3}}}{3!}\frac{{{\partial }^{3}}B}{\partial {{X}^{3}}}+\cdots.
\end{equation}
Therefore, by using Eq.~\eqref{eq:Taylor}, we rewrite Eqs.~\eqref{eq:EqMo_simplyfied} as follows~\cite{Deng2017,Deng2019}:
\begin{eqnarray}  \label{eq:EqMo_continuous}
    \frac{{{\partial }^{2}}{{B}_{u}}}{\partial {{X}^{2}}} &=& \frac{1}{a\left\{ \left( {{{V}^{2}}}/{c_{0,u}^{2}}\; \right)-1 \right\}}{{B}_{\theta }}\frac{\partial {{B}_{\theta }}}{\partial X} \\
    \kappa \frac{{{\partial }^{2}}{{B}_{\theta }}}{\partial {{X}^{2}}} &=& -6{{K}_{\theta }}{{B}_{\theta }}-2\left( 1+12{{{{C}'}}_{4}} \right)B_{\theta }^{3}-96{{{C}'}_{6}}B_{\theta }^{5} -4a{{B}_{\theta }}\frac{\partial {{B}_{u}}}{\partial X}+\frac{2a}{3}B_{\theta }^{3}\frac{\partial {{B}_{u}}}{\partial X} \label{eq:EqMo_continuous2}
\end{eqnarray}
where
\begin{equation} \label{eq:kappa}
    \kappa ={{a}^{2}}\left( \frac{{{V}^{2}}}{c_{0,u}^{2}{{\gamma }^{2}}}+{{K}_{\theta }}-1 \right)
\end{equation}
Here, we retain the nonlinear terms only up to third order.

Finally, following Ref.~\cite{Deng2017}, we integrate the first equation, giving 
\begin{equation} \label{eq:integration}
    \frac{\partial {{B}_{u}}}{\partial X}=\frac{1}{2a\left\{ \left( {{{V}^{2}}}/{c_{0,u}^{2}}\; \right)-1 \right\}}B_{\theta }^{2}+C,
\end{equation}
where $C$ is an integration constant.
In this study, we assume $C=0$ because our interest is traveling waves, i.e., solitons, with a finite temporal support.
By substituting Eq.~\eqref{eq:integration} into Eq.~\eqref{eq:EqMo_continuous2}, we get
\begin{equation} \label{eq:phi6}
    \frac{{{\partial }^{2}}{{B}_{\theta }}}{\partial {{X}^{2}}}={{\eta }_{1}}{{B}_{\theta }}-{{\eta }_{3}}B_{\theta }^{3}+{{\eta }_{5}}B_{\theta }^{5}
\end{equation}
where
\begin{eqnarray} 
    {{\eta }_{1}} &=& -\frac{6{{K}_{\theta }}}{\kappa }, \label{eq:eta1}\\
    {{\eta }_{3}} &=& \frac{2}{\kappa }\left\{ 1+12{{{{C}'}}_{4}}+\frac{1}{\left( {{{V}^{2}}}/{c_{0,u}^{2}}\; \right)-1} \right\}, \label{eq:eta3} \\
    {{\eta }_{5}} &=& \frac{1}{3\kappa }\left\{ \frac{1}{\left( {{{V}^{2}}}/{c_{0,u}^{2}}\; \right)-1}-288{{{{C}'}}_{6}} \right\}. \label{eq:eta5}
\end{eqnarray}
This is the Klein-Gordon equation with power law nonlinearities. 
This (static) equation of motion that can be obtained from a $\varphi^6$ model admits the following (nontopological) soliton solution \cite{Khare2008}:
\begin{equation} \label{eq:soliton}
    B_{\theta} = \frac{A_{\theta} \text{sech} (\sqrt{\eta_1}X)}{\sqrt{1-D \tanh^2 (\sqrt{\eta_1}X)}},
\end{equation}
where
\begin{eqnarray} \label{eq:coeff_soliton}
    A_{\theta}^2 &=& 2\eta_1(1+D)/\eta_3, \\
    (1+D)^2/D &=& 3\eta_3^2/(4\eta_1 \eta_5).
\end{eqnarray}

In Fig.~\ref{fig:phi6}(b), by using Eq.~\eqref{eq:soliton}, we plot the soliton solution as well as numerically measured a rotational angle profile of a solitary wave (Note 4). 
Here, the speed of the soliton is $V_s/c_0=0.49$, which is obtained from the simulations, and we use this wave speed to plot the analytical wave form. 
This analytical expression demonstrates good agreement with the numerical simulations. 

\begin{figure}[htbp]
    \centerline{ \includegraphics[width=1.0\textwidth]{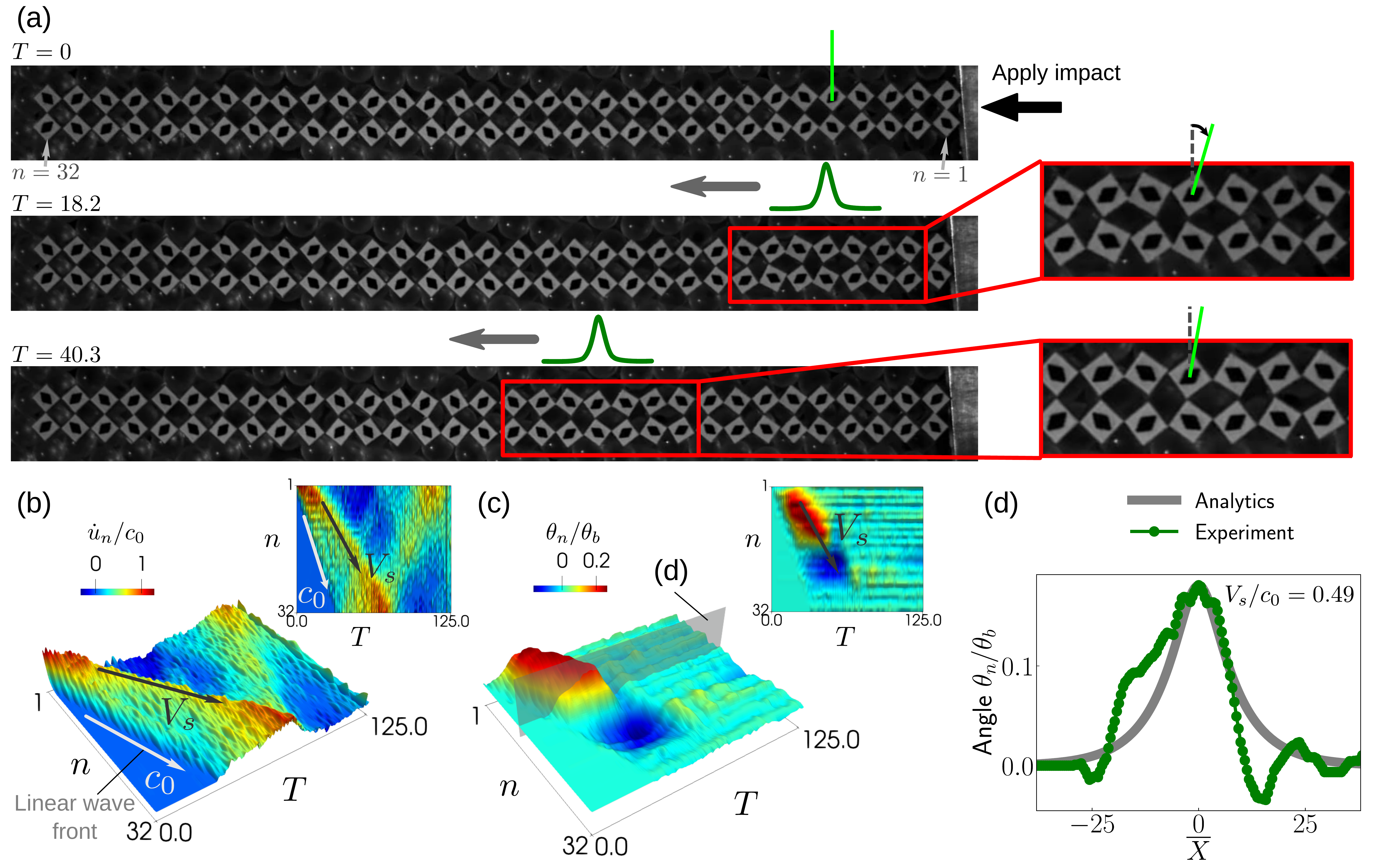}}
    \caption{ Comparison between the experimentally measured wavepacket and the (nontopological) soliton solution from the Klein-Gordon equation. (a) Snapshots of the experiment at $T=0$, 18.2, and 40.3. Here, $T=t c_0/a$. Note that $n=1$ corresponds to the right end of the chain to which we apply the impact input. The insets shows the magnified view of the generated wavepacket. The space-time evolution of the experimentally measured (a) velocity ($\dot{u}_n$) and (c) angle ($\theta_n$) wave propagation. (d) We plot the wave form of $n=13$ from the the experiment (green markers), which corresponds to the vertical (transparent) plane in panel (c), as well as the analytical solution (grey). Here, $\overline{X}=n-T$.}
    \label{fig:SI_single_soliton_exp}
\end{figure}

In addition to the comparison between the numerical simulation and the analytical soliton solution, we also demonstrate the propagation of a single vector soliton experimentally, and its wave form is compared with the analytical expression.
In the experiment, we apply the impact input only to the right end of the chain composed of 32 columns with $\theta^{(0)} = 0$ (see Fig.~\ref{fig:SI_single_soliton_exp}(a)).
By extracting translational and rotational motions of each rotating square element, we plot the measured velocities and angles as a function of time and space as shown in Fig.~\ref{fig:SI_single_soliton_exp}(b) and (c).
Our experiment shows two distinctive wave fronts; one is the linear wave (wave speed is bounded by the sound speed $c_0$) and the other is the localized wave packet.
Here, the wave speed of the experimentally measured wave packet is subsonic ($V_s/c_0=0.49$), and we confirm the subsonic nature of a vector soliton propagating in our system.
To conduct a more thorough analysis of this subsonically propagating wave packet, we extract its angle wave form and compare it with the analytical prediction based on the (nontopological) soliton solution of the nonlinear Klein-Gordon equation (Fig.~\ref{fig:SI_single_soliton_exp}(d)).
Although the experimentally measured wave shape does not show a clear (bell-shaped) solitay wave form, our theoretical approach predicts qualitatively similar wave shape.\\

\begin{figure}[htbp]
    \centerline{ \includegraphics[width=1.0\textwidth]{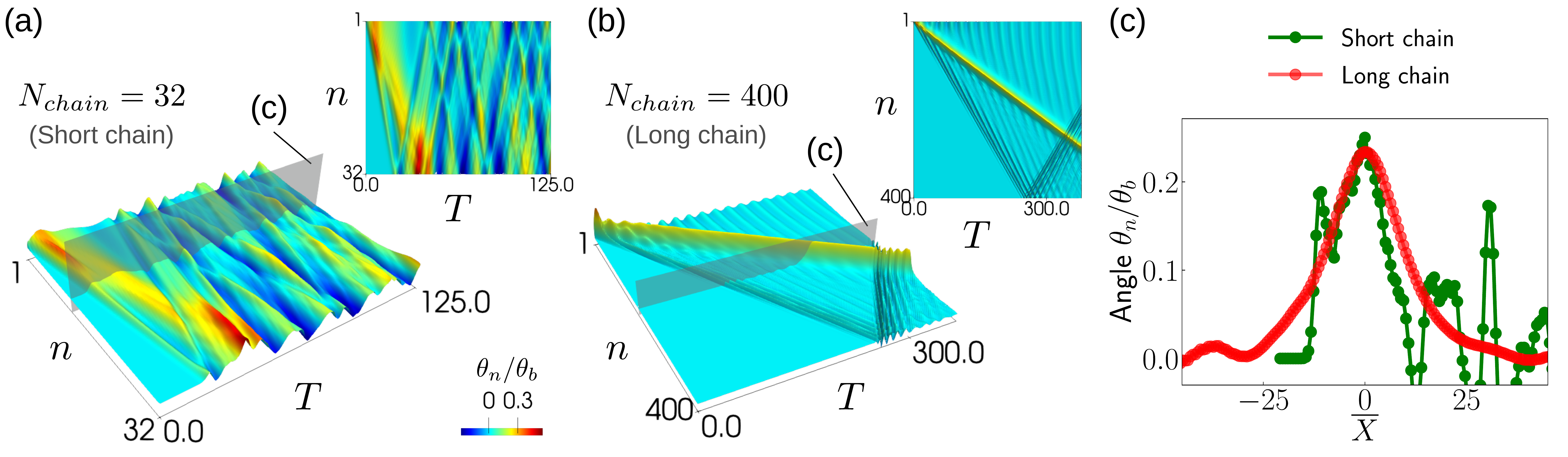}}
    \caption{ Spatiotemporal plots of angle wave propagation for two different chain length cases: (a) $N_{chain}=32$ and (b) $N_{chain}=400$. Note that we use the same initial conditions of $n=1$ for the two cases, $(\dot{u}_1, \dot{\theta}_1) = (1.0, 80.0)$. We extract the angle wave profiles of $n=13$ for $N_{chain}=32$ case (green markers) and $n=200$ for $N_{chain}=400$ case, and these two profiles are plotted in (c). Here, $T=t c_0/a$ and $\overline{X}=n-T$.}
    \label{fig:SI_single_soliton_sim}
\end{figure}

We further investigate such wave form of the vector soliton numerically by considering the effect of the chain length.
By setting nonzero initial conditions of the right end of the chain ($n=1$) as impact input, we analyze the wave propagation for two different chain length cases: $N_{chain}=32$ and 400.
Note that we use the identical initial conditions for those two cases so that the only difference is the chain length.
Figure~\ref{fig:SI_single_soliton_sim}(a) and (b) show the space-time contour plots of the angle variable for the shorter ($N_{chain}=32$) and longer ($N_{chain}=400$) chains, respectively.
For the shorter chain, although we observe a localized wave packet, its wave form shows a distorted shape because the impact input generates not only a localized pulse but also (small-amplitude) linear waves at the same time (see also the green markers in Fig.~\ref{fig:SI_single_soliton_sim}(c) for the wave form).
On the other hand, the longer chain allows the solitary wave to be distilled, which minimizes the effect of linear waves (and their reflected waves), and we observe the clear bell-shaped solitary wave (see the red markers in Fig.~\ref{fig:SI_single_soliton_sim}(c)).
Therefore, due to the limitations of our experiments (especially the chain length and nature of impact input), it is extremely challenging to observe the clear bell-shaped solitary wave experimentally.
However, based on the above discussion, we experimentally confirm the formation and propagation of a single solitary wave in our chain, which qualitatively agrees with our analytical prediction.

\clearpage
\section*{Supplementary  Note 6: \textbf{Collision behaviors}}

In addition to the collision case that triggers transition waves shown in the main text, we have performed tests with smaller amplitude pulses that do not form transition waves upon collision. We adopt the same dynamic testing setup shown in Fig.~\ref{fig:SI_dynamictest}, where the impact velocity and amplitude is controlled via the initial gap between the impactor and the structure. A smaller impactor-to-specimen gap would reduce the impact distance and therefore triggers a lower amplitude pulse. To compare the collision behaviors between two cases, we extract the rotational profile of each column and plot its angle wave form. 

In Figure \ref{fig:SI_failed collision}, we compare two cases side-by-side and observe two distinct collision cases: one leading to the formation of topological solitons and the other only generates dispersive waves. From the selected optical images in Figure \ref{fig:SI_failed collision}~(b), two types of nonlinear waves can be distinguished. At $t = 0~s$, two small amplitude initial pulses are sent from both ends and a transition wave front propagating at a significantly higher amplitude is later formed upon collision. 

For the scenario that does not induce transition waves, we generate two pulses where the signal starting from $n = 30$ is at a relatively lower amplitude. When the collision occurs in the proximity of $n = 16$, the pulses only cause slight perturbations upon impact and form an unstable nucleus, 
which later disintegrates into dispersive waves. The two pulses meet at approximately $t = 0.057 s$, and 3-4 neighboring units overcome the energy barrier, but later returns to the initial open state. Here, we demonstrate experimentally that collisions of waves can lead to two scenarios: 1. nucleation and propagation of transition waves, where the initial pulses overcome the energy barrier; 2. Disintegration of the primitive nucleus and later leads to the oscillation of dispersive waves. The experiments agree qualitatively with our theory that there exists an energy threshold for collisions of pulses to transform to another type of nonlinear wave (topological solitons) in a multistable lattice.
\begin{figure}[H]
\centerline{ \includegraphics[width=0.9\textwidth]{Figures_ResponseLetter/Collision_experiments.pdf}}
\caption{ Two collision cases comparison. Collision-induced phase transformation: (a) optical images taken with a high-speed camera
at times t = 0 s, 0.05 s, and 0.11 s and (b) Spatiotemporal surface plot of the rotational profile. Collision that does not induce topological solitons: (c) optical images taken with a high-speed camera
at times t = 0 s, 0.06 s, and 0.1 s and (d) Spatiotemporal surface plot of the rotational profile. }
\label{fig:SI_failed collision}
\end{figure}

\begin{figure*}[htbp]
    \centerline{ \includegraphics[width=1.0\textwidth]{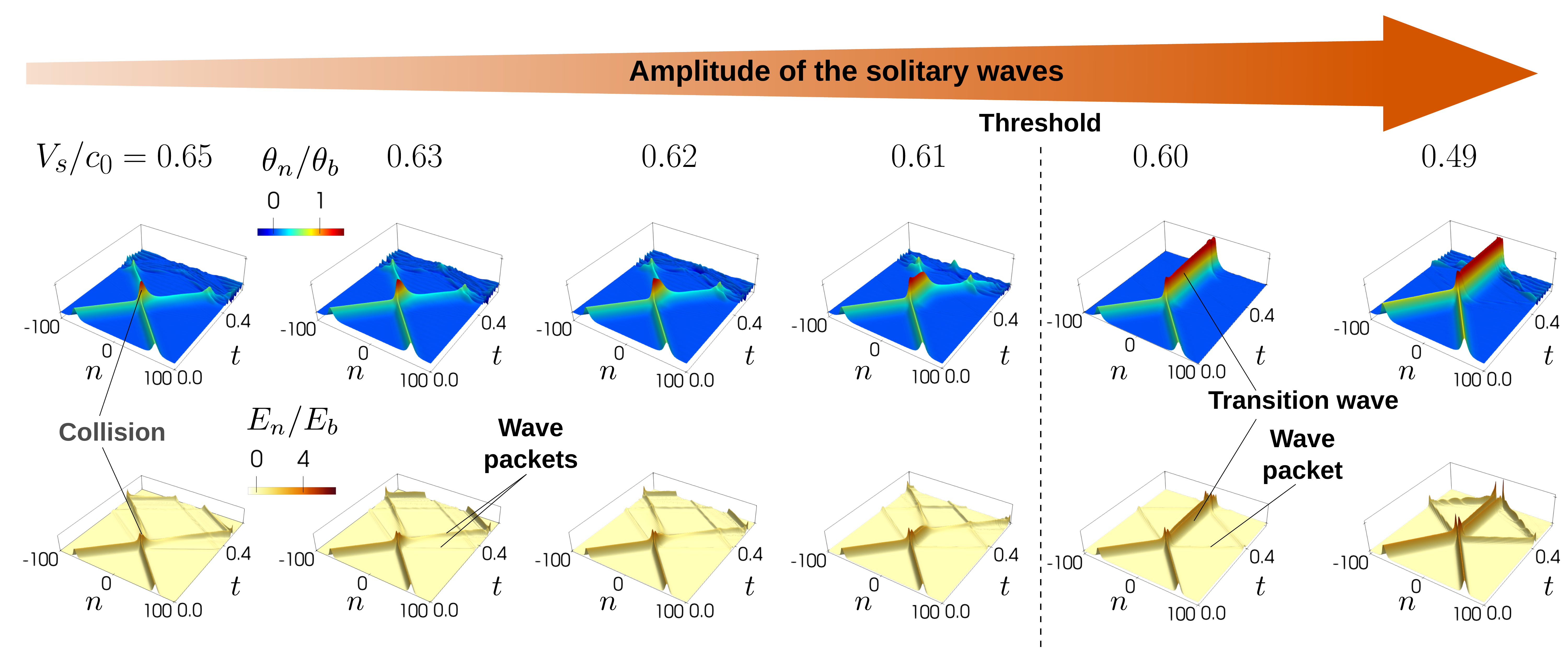}}
    \caption{Surface plots of (Upper) angle $\theta_n / \theta_b$ and (Lower) energy $E_n / E_b$ for different wave speeds ranging from $V_{s}/c_0=0.65$ (smaller amplitude) to $0.49$ (larger amplitude).} 
    \label{fig:wave_after_collision}
\end{figure*}

\begin{figure*}[htbp]
    \centerline{ \includegraphics[width=1.0\textwidth]{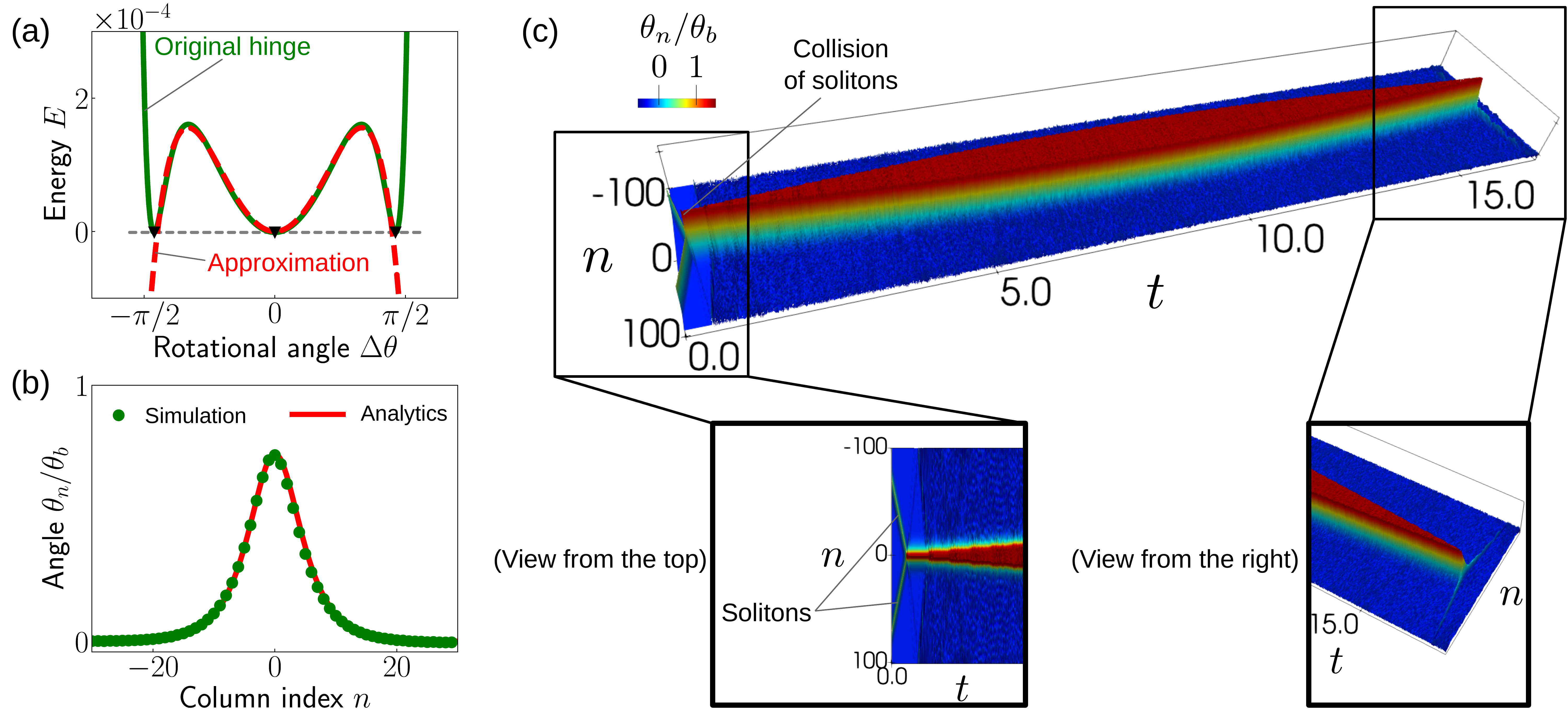}}
    \caption{(a) Energy landscape for $k_{\theta} = 5.04 \times 10^{-4}$ N m. The other numerical parameters are described in Table S1. All three energy minima denoted by the black triangles correspond to the same energy level. Note that the energy landscape discussed in the main text (see Fig.2(a)) shows that the energy level of the center energy well is higher than the other two wells. (b) The wave form obtained from the simulations is compared with the analytical solution for $V_{s}/c_0=0.52$. (c) The surface plot of rotational angles $\theta_n$ shows the collision of two vector solitons with $V_{s}/c_0=0.52$. Although the nucleation occurs once two vector solitons collide with each other, the propagation of transition waves is not well evolved compared with Fig.2(d) (main text) due to the altered energy landscape with no energy gap between different energy wells.} 
    \label{fig:wave_after_collision}
\end{figure*}

\begin{figure*}[htbp]
    \centerline{ \includegraphics[width=0.65\textwidth]{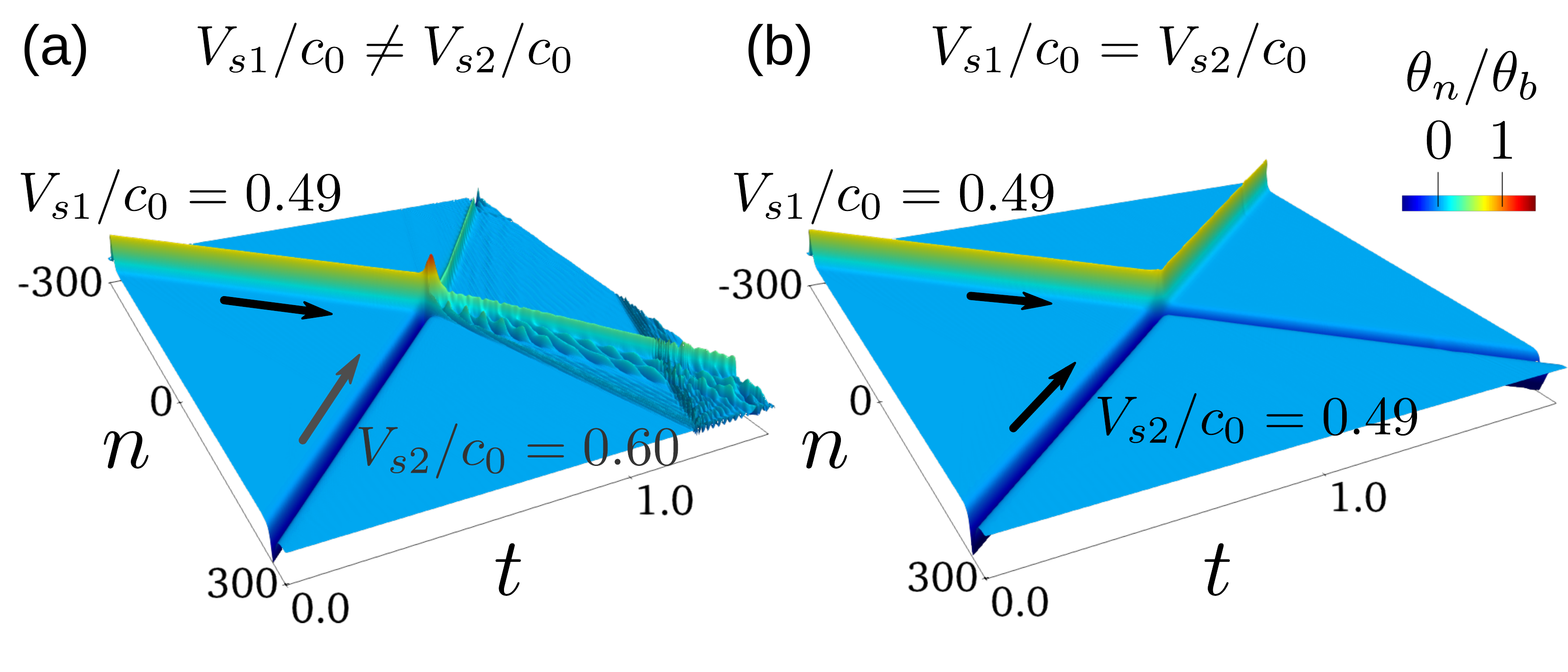}}
    \caption{Spatio-temporal plots of angle profiles show the head-on collision of two solitons with opposite rotation for (a) $V_{s1}/c_0=0.49$, $V_{s2}/c_0 =  0.60$ and (b) for $V_{s1}/c_0=V_{s2}/c_0 = 0.49$.
    } 
    \label{fig:headon_opposite}
\end{figure*}

\clearpage
\section*{Supplementary  Note 7: \textbf{Damping effect}}
\begin{figure}[htbp]
    \centerline{ \includegraphics[width=1.0\textwidth]{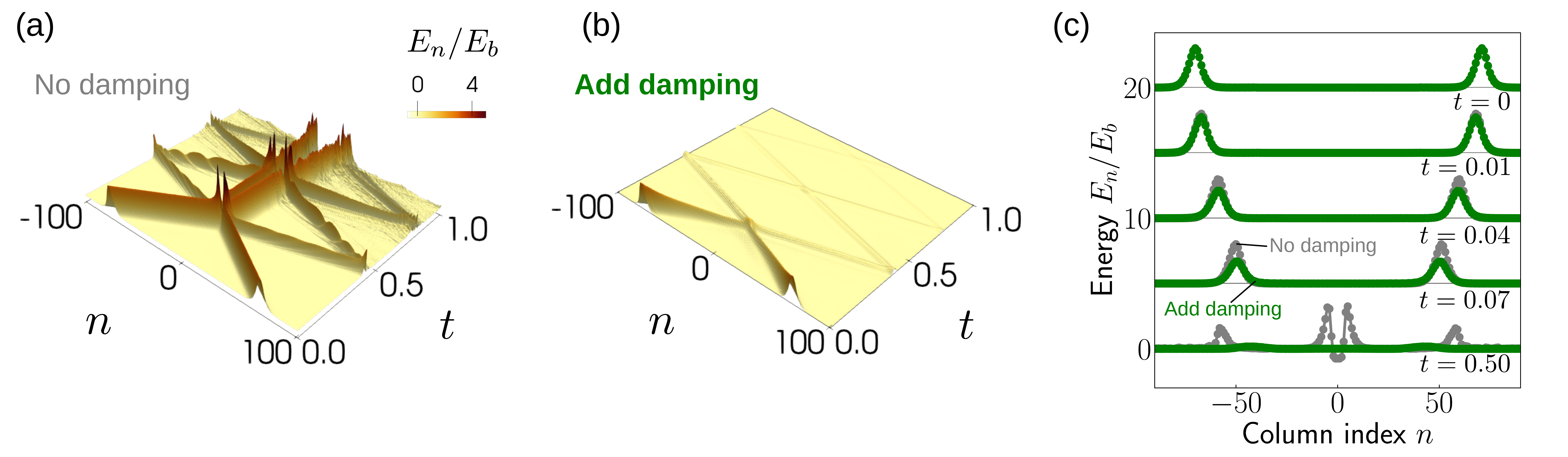}}
    \caption{The spatiotemporal surface plots of energy wave propagation in a chain (a) without damping and (b) with damping terms. The chain is composed of 202 columns. The wave speed of the initial vectors is $V/c_0=0.49$. (c) Wave forms at $t=0$ s, 0.01 s, 0.04 s, 0.07 s, and 0.50 s are plotted for the no damping case (grey) and damped case (green). Energy wave forms are offset in the vertical direction to ease visualization.}
    \label{fig:SI_damping_effect}
\end{figure}

Although we mainly investigate the nucleation of transition waves in our system without dissipation, we also consider the effect of damping by adding simple damping terms in the equations of motion as follows:
\begin{eqnarray}  \label{eq:EqMo_with_damping}
    {{M}_{n}}{{\ddot{u}}_{n}} &+& \nu_u \dot{u}_n = {{k}_{u}}\left( {{u}_{n+1}}+{{u}_{n-1}}-2{{u}_{n}} \right) \nonumber \\
    &+& {{k}_{u}}l\left\{ \cos \left( {{\theta }_{n-1}}+{{\theta }^{(0)}} \right)-\cos \left( {{\theta }_{n+1}}+{{\theta }^{(0)}} \right) \right\} \\
    {{J}_{n}}{{\ddot{\theta }}_{n}} &+& \nu_{\theta} \dot{\theta}_n = -{{k}_{\theta }}\left( {{\theta }_{n+1}}+{{\theta }_{n-1}}+4{{\theta }_{n}}+6{{\theta }^{(0)}}-6\theta _{Lin} \right) \nonumber \\
    &-& {{k}_{u}}l\sin \left( {{\theta }_{n}}+{{\theta }^{(0)}} \right)\left( {{u}_{n+1}}-{{u}_{n-1}} \right) \nonumber \\
    &+& {{k}_{u}}{{l}^{2}}\cos \left( {{\theta }_{n}}+{{\theta }^{(0)}} \right) \nonumber \\
    && \left\{ \sin \left( {{\theta }_{n+1}}+{{\theta }^{(0)}} \right)+\sin \left( {{\theta }_{n-1}}+{{\theta }^{(0)}} \right) \right\} \nonumber \\
    &-& {{T}_{Morse}}\left( \Delta {{\theta }_{n,1}} \right)-{{T}_{Morse}}\left( \Delta {{\theta }_{n-1,1}} \right) \nonumber \\
    &-& {{T}_{Morse}}\left( \Delta {{\theta }_{n,2}} \right)
\end{eqnarray}
where $\nu_u$ and $\nu_{\theta}$ are damping coefficients for translational and rotational motions, respectively.
Based on the previous study [36], we use $\nu_u=3.0$ and $\nu_{\theta}=136.7$ for this analysis.
Figure~\ref{fig:SI_damping_effect} shows the energy profiles for a chain composed of 202 columns with and without damping terms. 
Here, the initial wave speed of two vector solitons is $V(t=0)/c_0=0.49$ for both cases.
The system without damping shows the nucleation and propagation of transition waves after the collision (Fig.~\ref{fig:SI_damping_effect}(a)).
On the other hand, as the vector solitons propagate in the damped system, the waves lose energy, and the nucleation does not take place due to the insufficient amount of energy carried by the solitons (Fig.~\ref{fig:SI_damping_effect}(b)).
Figure~\ref{fig:SI_damping_effect}(c) shows the comparison between these two cases (with and without damping) in terms of the energy waveform.
The damped chain (green markers in the figure) exhibits a significant decrease of energy before the collision.

\begin{figure}[htbp]
    \centerline{ \includegraphics[width=1.0\textwidth]{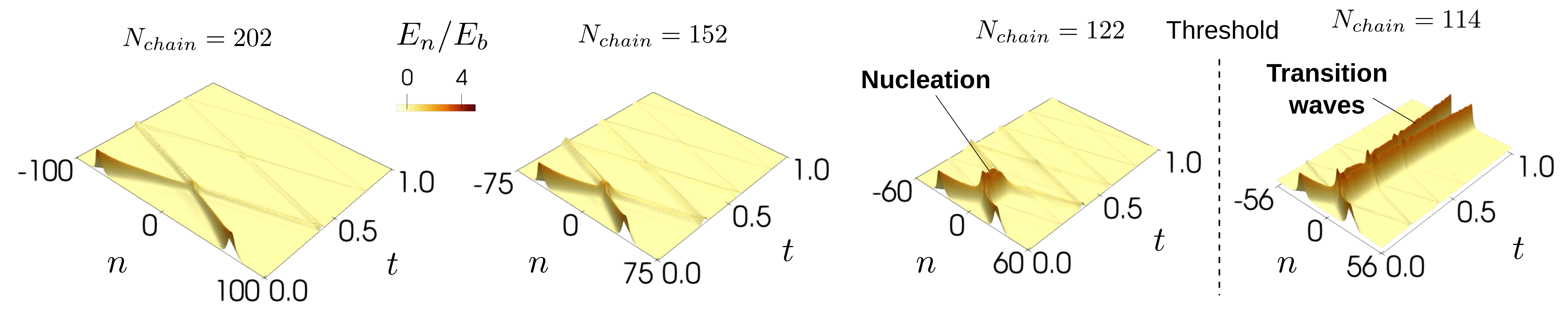}}
    \caption{The spatiotemporal surface plots of energy for various chain length cases: $N_{chain}=202$ (Left), 152, 122, and 114 (Right). The wave speed of the initial vector solitons is $V/c_0=0.49$.}
    \label{fig:SI_damping_chain_length}
\end{figure}

Then, we consider various chain length cases ($N_{chain}$) to identify the optimal length for triggering transition waves in the damped system.
Figure~\ref{fig:SI_damping_chain_length} shows different collision behaviors as we change the chain length ($N_{chain}=114, 122, 152, 202$).
If the chain is longer, the nucleation is not observed as we discuss above.
However, as we decrease the chain length, a decent amount of energy can be transported to the collision site, which triggers  nucleation.
In particular, the solitary waves lose less energy as they propagate in a shorter chain (e.g., $N_{chain}=114$), and propagation of transition waves can be observed.

\newpage